\documentclass[fleqn,usenatbib]{mnras}

\usepackage{newtxtext,newtxmath}
\usepackage[T1]{fontenc}

\DeclareRobustCommand{\VAN}[3]{#2}
\let\VANthebibliography\thebibliography
\def\thebibliography{\DeclareRobustCommand{\VAN}[3]{##3}\VANthebibliography}

\usepackage{graphicx}	% Including figure files
\usepackage{amsmath}	% Advanced maths commands
\newcommand{\beor}{\texttt{BayesEoR}}
\newcommand{\pyuvsim}{\texttt{pyuvsim}}
\newcommand{\figref}[1]{Figure \ref{#1}}
\newcommand{\hpx}{\texttt{HEALPix}}

\newcommand{\bs}{\ensuremath{\boldsymbol}}
\newcommand{\dvec}{\ensuremath{\boldsymbol{d}}}
\newcommand{\mvec}{\ensuremath{\boldsymbol{m}}}
\newcommand{\nvec}{\ensuremath{\boldsymbol{n}}}
\newcommand{\avec}{\ensuremath{\boldsymbol{a}}}
\newcommand{\qvec}{\ensuremath{\boldsymbol{q}}}
\newcommand{\phivec}{\ensuremath{\boldsymbol{\varphi}}}
\newcommand{\dbar}{\ensuremath{\bar{\dvec}}}

\newcommand{\Finvm}{\ensuremath{\mathbf{F}^{-1}}}
\newcommand{\Fprime}{\ensuremath{\mathbf{F}'}}
\newcommand{\Fz}{\ensuremath{\mathbf{F}_z}}
\newcommand{\Qz}{\ensuremath{\mathbf{Q}_z}}
\newcommand{\N}{\ensuremath{\mathbf{N}}}
\newcommand{\Ninv}{\ensuremath{\mathbf{N}^{-1}}}
\newcommand{\Psii}{\ensuremath{\bs{\Psi}^{-1}}}
\newcommand{\Sigmai}{\ensuremath{\bs{\Sigma}^{-1}}}
\newcommand{\T}{\ensuremath{\mathbf{T}}}
\newcommand{\rmd}{\ensuremath{\rm{d}}}

\newcommand{\freq}{\ensuremath{f}}
\newcommand{\enfreq}{\ensuremath{N_\freq}}
\newcommand{\enu}{\ensuremath{N_u}}
\newcommand{\env}{\ensuremath{N_v}}
\newcommand{\eneta}{\ensuremath{N_\eta}}
\newcommand{\enq}{\ensuremath{N_q}}
\newcommand{\ent}{\ensuremath{N_t}}
\newcommand{\enbls}{\ensuremath{N_{\rm bls}}}
\newcommand{\enmodel}{\ensuremath{N_{\rm model}}}
\newcommand{\envis}{\ensuremath{N_{\rm vis}}}

\title[All Sky Modelling Requirements with BayesEoR]{All Sky Modelling Requirements for Bayesian 21 cm Power Spectrum Estimation with BayesEoR}

\author[J. Burba et al.]{
Jacob Burba,$^{1,2}$\thanks{E-mail: jacob.burba@manchester.ac.uk (JB)}
Peter Sims,$^{3,4}$
Jonathan C. Pober$^{2}$
\\
$^{1}$Department of Physics and Astronomy, University of Manchester, Manchester, UK\\
$^{2}$Department of Physics, Brown University, Providence, RI, USA\\
$^3$McGill Space Institute, McGill University, 3550 University Street, Montreal, QC H3A 2A7, Canada \\
$^4$Department of Physics, McGill University, 3600 University Street, Montreal, QC H3A 2T8, Canada \\
}

\date{Accepted XXX. Received YYY; in original form ZZZ}

\pubyear{2022}

\begin{document}
\label{firstpage}
\pagerange{\pageref{firstpage}--\pageref{lastpage}}
\maketitle

\begin{abstract}
We present a comprehensive simulation-based study of the BayesEoR code for 21 cm power spectrum recovery when analytically marginalizing over foreground parameters.  To account for covariance between the 21 cm signal and contaminating foreground emission, BayesEoR jointly constructs models for both signals within a Bayesian framework.  Due to computational constraints, the forward model is constructed using a restricted field-of-view (FoV) in the image domain.  When the only EoR contaminants are noise and foregrounds, we demonstrate that BayesEoR can accurately recover the 21 cm power spectrum when the component of sky emission outside this forward-modelled region is downweighted by the beam at the level of the dynamic range between the foreground and 21 cm signals. However, when all-sky foreground emission is included along with a realistic instrument primary beam with sidelobes above this threshold extending to the horizon, the recovered power spectrum is contaminated by unmodelled sky emission outside the restricted FoV model.  Expanding the combined cosmological and foreground model to cover the whole sky is computationally prohibitive.  To address this, we present a modified version of BayesEoR that allows for an all-sky foreground model, while the modelled 21 cm signal remains only within the primary FoV of the telescope.  With this modification, it will be feasible to run an all-sky BayesEoR analysis on a sizeable compute cluster.  We also discuss several future directions for further reducing the need to model all-sky foregrounds, including wide-field foreground subtraction, an image-domain likelihood utilizing a tapering function, and instrument primary beam design.

% It should be a single paragraph not more than 250 words (200 words for Letters).
% No references should appear in the abstract.
\end{abstract}

\begin{keywords}
methods: data analysis -- dark ages, reionization, first stars -- techniques: interferometric
\end{keywords}

%%%%%%%%%%%%%%%%%%%%%%%%%%%%%%%%%%%%%%%%%%%%%%%%%%

%%%%%%%%%%%%%%%%% BODY OF PAPER %%%%%%%%%%%%%%%%%%

\defcitealias{S16}{S16}
\defcitealias{S19a}{S19a}
\defcitealias{S19b}{S19b}

\section{Introduction}

The Epoch of Reionization (EoR) marks the period in the universe's history where the neutral hydrogen (HI) in the intergalactic medium became ionized by the first luminous sources.  This epoch contains a wealth of information about the early universe and the structure and form of the first stars and galaxies \citep{furlanetto_oh_briggs, morales_wyithe, pritchard_loeb, zaroubi, barkana, mesinger, liu-shaw}.  Observing this distant epoch, however, brings with it a unique set of challenges.

Current interferometric experiments like the Murchison Widefield Array (MWA, \cite{mwa}), Low Frequency ARray (LoFAR, \cite{lofar}), and the Hydrogen Epoch of Reionization Array (HERA, \cite{hera}) are attempting to observe this HI signal during the EoR via the redshifted 21 cm line.  While the eventual goal of these experiments is to perform 21 cm tomography, current efforts for a first detection are focused on estimating the power spectrum from 21 cm data.  Estimation of the 21 cm EoR power spectrum has proven difficult, however, due to the presence of bright, contaminating sources in between us and the cosmological signal referred to as ``foregrounds'' (FGs) \citep{lofar-limits, mwa-limits, hera-uplims}.

\citet{liu-shaw} provides a summary of current techniques for dealing with FGs during power spectrum estimation from 21 cm interferometric data.  These FG mitigation techniques can be roughly divided into two main categories: subtraction and avoidance.  In the former, a FG model is constructed and subtracted from the data prior to EoR power spectrum estimation. In the latter, power spectrum modes dominated by FGs (and/or instrumental effects) are avoided entirely when estimating the EoR power spectrum.  An ideal analysis, however, would be capable of jointly modelling the EoR and FGs to properly account for the covariance between the two observed signals.  This joint analysis of EoR and FGs is possible via a Bayesian framework and is a key advantage of our code \beor\footnote{\url{https://github.com/PSims/BayesEoR}}.

Bayesian studies of EoR and FG signal separation have grown in popularity in the 21 cm community in recent years.  \citet{ghosh15} used a Bayesian framework to compute a maximum a posteriori (MAP) image cube from simulated LOFAR visibilities containing mock EoR and diffuse FG signals.  They then employed a generalized morphological component analysis (GMCA) to model and subtract the FG signal from the MAP image cube, the residuals of which produced unbiased EoR power spectrum spectrum estimates largely consistent with the input EoR power spectrum.  This approach, however, ultimately risks over (under) subtracting the FG signal in the data if an overly complex (simplistic) FG model is used.  If the EoR and FG signals are correlated in the visibilities, as is the case for real data, this approach can also result in incorrect uncertainty estimates on the EoR power spectrum.  In their more recent work, \citet{ghosh20} employ a Gaussian process regression (GPR) technique within a Bayesian framework to produce a set of FG model posteriors from observed HERA data. The examination of the residual data from which the EoR power spectrum would be estimated, however, was left as future work.  Further studies using simulated datasets with known inputs are required to determine the efficacy of this approach in regards to alteration of the 21 cm signal.  Similar to our approach, \citet{zhang16} employ a Bayesian framework and jointly model the EoR and FG signals in the data.  However, the large dynamic range between the EoR and FG signals compounded with the mode mixing introduced via observation with an interferometer make this joint model challenging (as is the case for our analysis).  The latter effect modulates the intrinsic smoothness of the FG signal with the spectral structure of the instrument, inducing correlations between the observed EoR and FG signals.
A viable approach for real data must account for the frequency dependence of both the sampling of the interferometer in the $uv$-plane and the point spread function (PSF).  The work performed in \citet{zhang16} did not account for the spectral nature of the instrument and thus requires further modification for use on real data.

Fortunately, the drawbacks of the above approaches can be overcome via the use of our Bayesian approach to power spectrum estimation.  By jointly modelling the EoR and FG signals and forward modelling the instrument, we can account for the observed covariance between the EoR and FG signals.  This joint fitting also ensures that uncertainties on all model components are encapsulated in the final power spectrum estimates.  Because we eventually wish to perform astrophysical parameter inference (e.g. \citet{21cmmc}, \citet{hera-theory}), robust uncertainties on power spectrum estimates will be essential to avoid biased astrophysical parameter inferences.  Additionally, our code is flexible enough to go beyond the power spectrum, allowing for the introduction of non-Gaussian priors on the EoR signal (see section 3.1 of \citet{S19b}) and affords the use of Bayesian model selection to compare physically motivated signal parametrizations.

\citet{S16}, \citet{S19a}, and \citet{S19b} (hereafter \citetalias{S16}, \citetalias{S19a}, and \citetalias{S19b}, respectively) outline the mathematics and provide demonstrations using simulated datasets of our Bayesian approach to 21 cm interferometric power spectrum estimation.  \beor\ estimates the power spectrum from a set of visibilities by forward modelling the instrument.  This forward model is comprised of a series of Fourier transforms that takes a set of $k$-space amplitudes to a set of instrumentally sampled visibilities (described in more detail in section \ref{sec:beor-model}).  To properly model a set of visibilities, one must account for the primary beam response of the antennas.  As such, our forward model involves an intermediate transformation to the image domain where we multiply the intrinsic sky by the instrument beam.  We then transform the product of sky times beam to form a set of model visibilities.

While including the entire sky horizon-to-horizon in the image domain model is in principle possible, doing so is computationally demanding.  Typically, only the sky inside the main lobe of the primary beam is used to forward model visibilities inside \beor.  All previous published works using \beor\ thus used input visibilities simulated from a rectangular subset of sky which matched exactly with the internal image domain model.  In this way, the input data to \beor\ exclusively contain information about the sky on the scales accessible to the forward model.
However, real instruments are in general sensitive to emission from horizon to horizon due to the primary beam response of the antennas\footnote{The wide field nature of these instruments also makes them particularly sensitive to polarization leakage which maximizes on the horizon \citep{asad15, asad16, asad18}.} \citep{nfbeams, lofar-beams, mwa-beams, ska-low-beam-2016, ska-low-beam-2017}.
Visibilities from a real instrument thus contain information about the entire sky.  In theory, if the beam sufficiently downweights sky emission far from field-center, high zenith angle FGs should contribute marginally to the observed visibilities.  In this case, a restricted image domain model in \beor\ should be sufficient for forward modelling these visibilities.  Conversely, if the beam insufficiently downweights sky emission far from field-center, there will be information about the sky in the observed visibilities that a restricted image domain model cannot reproduce.  A key question then is how well we can model a set of visibilities from a realistic instrument using a restricted image domain model.  To address this question, we applied \beor\ to a set of detailed simulations using various primary beam patterns that observe either a subset of the sky or the entire sky. 

While exploring the full parameter space of our model is possible, it requires the use of sampling algorithms designed to efficiently explore large parameter spaces.  Marginalization, on the other hand, has the significant advantage of greatly reducing the size of the parameter space.  This facilitates powerful Bayesian model selection techniques for FG model optimization and signal detection quantification via nested sampling that are unavailable with alternate sampling techniques (see e.g.\ \citetalias{S19b}).
It is for these reasons that we exclusively used the marginalized form of the posterior presented in \citetalias{S16} for this work.
While we discuss the computational trade-offs of these techniques in more detail in section \ref{sec:real-data-techniques}, we mention this here to better frame the subsequent discussion.

In this work, we assume that the following effects have been dealt with accurately in pre-processing of the data, i.e.\ prior to power spectrum estimation with BayesEoR: calibration, polarization leakage, ionospheric effects, radio frequency interference, and per-antenna beams.  Ultimately, we intend to encompass these effects in a comprehensive Bayesian framework.  Bayesian power spectrum estimation with \beor\ and calibration with \texttt{BayesCal} \citep{bayescal-i, bayescal-ii} represent the first two steps in this larger framework.  We leave a discussion of this larger framework to future work, however.
Here, the only EoR contaminants in our simulated datasets are FGs and noise.  This work is thus intended to solely demonstrate the impact of incomplete foreground modelling during power spectrum estimation.

The remainder of this paper is laid out as follows.  In section \ref{sec:beor-ps-estimation}, we provide a high level overview of the mathematical machinery behind \beor\ and describe the modifications to the analysis from previous works.  In section \ref{sec:simulations}, we describe our simulated datasets used as the input data to \beor.  In sections \ref{sec:results} and \ref{sec:discussion}, we present and discuss, respectively, the results of our power spectrum analyses.
In section \ref{sec:real-data-techniques}, we present the computational constraints associated with sampling from the marginalized posterior and techniques for overcoming them.
Lastly, in section \ref{sec:conclusion} we summarize our results and provide future directions.

\section{Power Spectrum Estimation}\label{sec:beor-ps-estimation}

\beor\ utilizes Bayesian inference to estimate a set of parameters, $\Theta$, from a model, $\mvec(\Theta)$, given a set of data (visibilities), $\dvec$.
For consistency with \citetalias{S16} and \citetalias{S19a}, henceforth, our model and data vectors will refer to the concatenation of the real and imaginary components rather than a single vector of complex values.  If we have $\envis$ visibilities in our data, our model and data vectors will then have a length of $2\envis$.  For this reason, in the following equations, we will use the transpose symbol $T$ in place of the Hermitian conjugate symbol $\dagger$.
Unless otherwise specified, the terms ``data'' and ``visibilities'' will be used interchangeably.  In the following subsections, we present a high level overview of the model and posterior utilized by \beor.  For more detail, we refer the reader to \citetalias{S16}, \citetalias{S19a}, and \citetalias{S19b}.

\subsection{Model}\label{sec:beor-model}

The model is constructed via a series of discrete Fourier transforms (DFTs) applied to a rectilinear, three-dimensional grid in $k$-space, ($k_x$, $k_y$, $k_z$).  Using the relations
\begin{align}
    k_x &= \frac{2\pi u}{D_m}   \nonumber\\
    k_y &= \frac{2\pi v}{D_m}   \nonumber\\
    k_z &= \frac{2\pi H_0\freq_{21}E(z)}{c(1+z)^2}\eta
\end{align}
we can equivalently define this $k$-cube in terms of $(u, v, \eta)$ coordinates where $u$ and $v$ are the $uv$-coordinates sampled by a given baseline and $\eta$ is the Fourier dual of frequency.  In the equations above, $z$ is the redshift of the observation, $D_m$ is the transverse comoving distance to this redshift, $H_0$ is the Hubble constant, $E(z)$ is the dimensionless Hubble parameter, $\freq_{21}$ is the frequency of the 21 cm line emission, and $c$ is the speed of light.
We define the set of complex $(u, v, \eta)$ amplitudes corresponding to the 21 cm EoR signal in our model as \avec.
To model FG spectral structure in the input visibilities on scales larger than the bandwidth, we also define the set of Large Spectral Scale Model (LSSM) coefficients per model $(u, v)$ as \qvec.  Using the combination of \avec\ (EoR) and \qvec\ (FGs), we can then define our model visibilities (\mvec, see \citetalias{S16} for details) via a series of DFTs as
\begin{equation}
    \mvec = \Finvm \Fprime (\Fz\avec + \Qz\qvec)
    \label{eq:model-vector}
\end{equation}
\Fz\ is a one-dimensional DFT matrix that samples the data on scales less than or equal to the bandwidth.  \Qz\ is a matrix containing the LSSM basis vectors which sample the data on scales larger than the bandwidth.  For the LSSM, we use the constant plus double power law (CPDPL) model from \citetalias{S19b} defined by
\begin{equation}
    \Qz\qvec = q_0 + q_1\left(\frac{\bs{\freq}}{\freq_0}\right)^{b_1} + q_2\left(\frac{\bs{\freq}}{\freq_0}\right)^{b_2}
    \label{eq:cpdpl}
\end{equation}
where $\bs{\freq}$ is a vector of frequencies in the data and model and $b_1=\left< \beta \right>_{\rm{GDSE}}=-2.63$ and $b_2=\left< \beta \right>_{\rm{EGS}}=-2.82$ are the mean brightness temperature spectral indices for galactic diffuse synchrotron and extragalactic source FG emission \citepalias{S19b}.
\Fprime\ is a two-dimensional DFT matrix relating the model $uv$-plane to the image domain.  Lastly, \Finvm\ is a matrix relating the image domain to the instrumentally sampled, beam convolved model visibilities.
Note that \Finvm\ contains all of the information about the instrument being modelled, e.g.\ the $uv$-sampling and primary beam.  The model therefore knows nothing about the instrument prior to the application of \Finvm, aside from the frequency axis.  The vectors \avec\ and \qvec\ therefore represent the intrinsic EoR and FG signals un-corrupted by the instrument.  In this way, the image domain representation of the model, $\Fprime(\Fz\avec + \Qz\qvec)$, similarly represents the brightness temperature of the sum of the intrinsic EoR and FG signals.
While the functions of \Fprime\ and \Finvm\ remain the same, they take on slightly different forms for this work due to a differing gridding scheme in the image domain.

\subsubsection{Modifications}\label{sec:mods}

The model $uv$-plane is constructed as a rectilinear grid in $(u, v)$ ($uv$-space).  In previous works, the image domain model was constructed as a complimentary rectilinear grid in $(l, m)$.  In this work, we have modified the image domain to use a \hpx\ \citep{healpix} grid in (RA, Dec) from which we can derive the local coordinates $(l, m, n)$ (image-space).

The motivations for the introduction of the \hpx\ grid in the image domain are two-fold.  First, a \hpx\ grid naturally captures curved sky effects, which are important to model for moderate to large FoV instruments.  With a narrow enough field of view (FoV), the flat and curved sky approaches will be comparable.  But, as discussed later, it is in some cases necessary to use a large FoV to accurately model data from a given instrument.  Second, the simulated visibilities used as the input mock data to \beor\ were obtained via \pyuvsim\ \citep{pyuvsim}\footnote{\url{https://github.com/RadioAstronomySoftwareGroup/pyuvsim}}, a high-precision visibility simulator which uses the \hpx\ gridding scheme to describe diffuse emission.

In generating a set of model visibilities, \beor\ performs its own internal visibility simulation\footnote{Note that the ``simulated'' visibilities from \Finvm\ are not used as the input data to \beor.  The outputs of \Finvm\ represent the model visibilities that are compared to the input data in the likelihood.  All input data in this work were simulated externally using \pyuvsim.}.  \Finvm\ takes a \hpx\ grid of frequency and time dependent sky $\times$ beam amplitudes and transforms them to a set of instrumentally sampled model visibilities.  It is vital that these model visibilities produced by \Finvm\ be accurate.  If we cannot properly model our data, we will be unable to recover the true power spectrum in the data.  The machinery powering \pyuvsim\ has been rigorously tested and shown to accurately reproduce a set of analytically viable visibility solutions \citep{lanman-murray-jacobs, validation}.  \pyuvsim\ therefore provides a robust reference against which we can compare the model visibilities from \Finvm.  For both diffuse signals used in this work (mock EoR and global sky model, described in section \ref{sec:sky-signals}), visibilities from \pyuvsim\ and \Finvm\ were found to agree to the same high level of precision when using a matching \hpx\ resolution.  Computing the fractional difference of the two sets of visibilities (1 - \beor\ / \pyuvsim) yielded $10^{-13}$ and $10^{-11}$ in amplitude and phase, respectively.  \Finvm\ is thus capable of producing highly accurate visibilities from a \hpx\ sky model.

Additional testing has been done using mock EoR only datasets where the nside value, which sets the resolution of the \hpx\ grid, in the data and model were independently varied.  A set of simulated datasets were simulated using nside values ranging from 32 - 1024.  For each simulated dataset, i.e.\ each input data nside, power spectra were estimated using model nside values of 32 - 512.  For all input data nside values, the recovered EoR power spectra for model nside values $\geq64$ were found to be consistent with the expected power spectrum amplitude at all $k$.  The model nside=32 power spectra, regardless of the nside of the input data, were biased low as the resolution of the image domain model did not satisfy the Nyquist-Shannon sampling theorem \citep{nyquist-shannon}.  If the maximum $|u|$ sampled by the model $uv$-plane is $|u|_{\rm max}$, Nyquist sampling the image domain requires a maximum pixel separation of $1/2|u|_{\rm max}$, i.e. having at least two pixels per minimum fringe wavelength $1/|u|_{\rm max}$.  The model $uv$-plane in these tests had $|u|_{\rm max} = 25.109$ rad$^{-1}$.  The required minimum pixel separation is thus $1/2|u|_{\rm max} \approx 0.02$ rad.  At nside$=32$, the \hpx\ pixel width is $\approx0.032$ rad $> 0.02$ rad.  For reference, at nside$=64$, where the recovered power spectrum estimates were consistent with expectation at all $k$ regardless of the input data nside, the \hpx\ pixel separation is $\approx0.016$ rad < $0.02$ rad.  The resolution of the image domain model therefore must be chosen to satisfy the Nyquist-Shannon sampling theorem.

\subsection{Posterior}

While the aforementioned modifications adjust the coordinates associated with the individual matrices that comprise the model, the form of the posterior remains unchanged.  Here, we present a brief overview of the final form of this posterior.  For a detailed derivation, we refer the reader to \citetalias{S19a}.

We begin by assuming that our data vector, $\dvec$, is a sum of signal, $\bs{s}$, and uncorrelated Gaussian random noise, \nvec.
Under this assumption, we can define the covariance matrix of the data as
\begin{equation}
    \mathbf{N} = \left< n_in_j^* \right> = \sigma_i^2\delta_{ij}
\end{equation}
where $<...>$ represents the expectation value and $\delta_{ij}$ is the Kronecker delta function.
While the assumption of a diagonal noise covariance matrix is fairly common in the literature \citep{ghosh15, zhang16, mertens18, lofar-limits, liu-shaw, dom}, we are interested in studies involving more general, dense noise covariance matrices.  However, we defer the exploration of the effects of data with non-diagonal noise covariance matrices to future work.

The simulated visibilities used here are noise free and represent the signal.  At run time, the data vector is formed by generating and adding noise to the input signal.  This noise is drawn from a Gaussian distribution with mean zero and a standard deviation equal to twice the standard deviation of the EoR only visibilities.  The signal to noise ratio (SNR) in visibility space of the EoR signal in all analyses is thus 0.5, chosen as a fiducial value that allows for clear detections of the EoR power spectrum at all scales in simulated datasets containing only the mock EoR signal.
This choice of noise allows us to write down a Gaussian likelihood function for the data model given by
\begin{align}
    \mathcal{L}(\avec, \qvec)
    \propto
    \frac{1}{\sqrt{\det(\N)}}\exp\left[ -\frac12\big(\dvec - \mvec(\avec, \qvec)\big)^T\Ninv\big(\dvec - \mvec(\avec, \qvec)\big) \right]
    \label{eq:likelihood}
\end{align}

Under the assumption that the redshifted 21 cm signal is homogeneous and isotropic, we can define the covariance matrix of the $k$-space coefficients \avec\ by
\begin{equation}
    \Psi_{ij} = \left< a(k_i) a^*(k_j) \right> = \varphi_i\delta_{ij}\ .
\end{equation}
Here, $\varphi_i$ represents the dimensionless power spectrum amplitude in the $i$-th spherically averaged $k$-bin in units of mK$^2$.  Given the joint probability distribution of the power spectrum and model
\begin{equation}
    \Pr(\phivec, \avec, \qvec | \dvec)
    \propto
    \Pr(\dvec|\avec, \qvec)\Pr(\avec | \phivec)\Pr(\phivec)\Pr(\qvec)
    \label{eq:joint-posterior-form}
\end{equation}
we can use our knowledge of the covariance of $\phivec$ to write that
\begin{equation}
    \Pr(\avec|\phivec)\Pr(\phivec)
    \propto
    \frac{1}{\sqrt{\det(\bs{\Psi})}}\exp\left[ -\frac12\avec^T\Psii\avec \right]
\end{equation}
Here, we have explicitly assumed that $\Pr(\phivec)=1/\varphi$, i.e. chosen a log-uniform prior on the power spectrum coefficients.
When extracting an upper limit from a given power spectral bin, however, a prior that is uniform in the amplitude is appropriate.  When using a uniform prior on $\phivec$, we instead have that
\begin{equation}
    \Pr(\avec|\phivec)\Pr(\phivec)
    \propto
    \frac{1}{\sqrt{\det(\bs{\Psi})}}\exp\left[ -\frac12\avec^T\Psii\avec \right]
    \prod_{s=1}^{N_s}\varphi_s
\end{equation}
where $N_s$ is the number of spherical power spectrum $k$-bins.
Consistent with previous works, we assume a uniform prior on the LSSM coefficients, i.e.\ $\Pr(\qvec)=1$.

If we further define $\T=\Finvm\Fprime(\Fz+\Qz)$, then we can define the quantity $\dbar=\T^T\Ninv\dvec$ which represents the projection of the covariance-weighted visibilities on the $k$-space grid of the model parameters.  After marginalizing over the signal coefficients \avec\ and \qvec\ (see \citetalias{S16} for details), the resulting joint posterior takes the form
\begin{align}
    \Pr(\bs{\varphi}|\bs{d})\Pr(\bs{\varphi})
        &\propto
            \frac{\det(\bs{\Sigma})^{-1/2}}{\sqrt{\det(\bs{\Psi})\det(\N)}} \nonumber\\
        &\times
            \exp\left[ -\frac12\left( \dvec^T\Ninv\dvec - \dbar^T\Sigmai\dbar \right) \right]
\end{align}
with $\bs\Sigma = \T^T\Ninv\T + \Psii$ the covariance matrix of \dbar.  When instead assuming a uniform prior on the power spectrum amplitudes $\bs{\varphi}$, the above posterior is multiplied by
\begin{equation}
    \Pr(\bs{\varphi}) \propto \prod_{s=1}^{N_s} \varphi_s
\end{equation}

\subsection{Model Parameters}\label{sec:model-params}

For clarity in subsequent discussions, we briefly define here some useful model parameters.  The $k$-space model in BayesEoR is constructed as a rectilinear 3D grid in ($k_x$, $k_y$, $k_z$), where $k_x$, $k_y$, and $k_z$ map to $u$, $v$, and $\eta$, respectively.  The shape of the $k$-space model is determined by the parameters \enu, \env, and \eneta\ which represent the number of sampled Fourier modes along the $k_x$, $k_y$, and $k_z$ axes, respectively.  The resolution of each axis is determined by the chosen field of view (FoV) (or bandwidth) along the corresponding axis in image-space.  Let us define FoV$_{\rm{RA}}$ and FoV$_{\rm{Dec}}$ as the fields of view in our image-space model along the right ascension (RA) and declination (Dec) axes.  Then, the spacing along the $u$ and $v$ axes are given by $\Delta u = \rm{FoV}_{\rm{RA}}^{-1}$ and $\Delta v = \rm{FoV}_{\rm{Dec}}^{-1}$, respectively.  Accordingly, for a dataset spanning a bandwidth $B$, we have that $\Delta\eta=B^{-1}$.

Choosing the right combination of these model parameters requires detailed knowledge of the instrument being modelled.  As previously mentioned, the chosen model FoV values along the RA and Dec axes determine the spacing between adjacent modes in the model $uv$-plane.  For a given FoV, \enu\ and \env\ must be chosen to fully encompass the baselines sampled by the instrument.  Additionally, a buffer must also be provided to account for the width of the aperture function.  As an example, let us consider an Airy disk corresponding to an antenna diameter of 14.6 m.
The aperture function of this beam pattern has an approximate width of $14.6 /\lambda$ wavelengths.  A minimum buffer of $0.5\cdot14.6 /\lambda$ wavelengths beyond the longest baseline in the data is thus necessary to fully capture the information in the visibilities.
To illustrate this, let us consider the baselines sampled by our chosen instrument model in the bottom panel of \figref{fig:inst-model}.  For a FoV along RA and Dec of $\sim19.4^\circ$, encompassing the longest baselines without consideration for the beam requires \enu$\,\geq19$.  When accounting for the beam, we instead require \enu$\,\geq23$.  If we choose to double the FoV of our model, we must also double \enu\ to fully encompass the information in the $uv$-plane.  This interplay between the FoV and \enu\ has important ramifications for the efficiency of \beor\ which will be discussed later.  Ultimately, the model FoV, \enu, and the aperture function must all be accounted for when choosing a set of model parameters.

\section{Simulations}\label{sec:simulations}

The data used here as the input to \beor\ are visibilities simulated with \pyuvsim\ \citep{pyuvsim}.  Common parameters among all simulations can be found in Table \ref{tab:simparams}.
For this choice of simulation parameters, and using the baselines from the lower panel of \figref{fig:inst-model}, we have $\envis=\enbls\cdot\enfreq\cdot\ent=38,760$ visibilities.  For reference, with a chosen FoV for the image domain model along RA and Dec of 19.4$^\circ$, as discussed in section \ref{sec:model-params}, we require $\enu=\env=23$ to fully encompass the baselines \figref{fig:inst-model}.  This results in $\enmodel=\enu\cdot\env\cdot(\enfreq+\enq)=21,160$ model parameters.
In the following subsections, we describe the array layout, sky signals, and beam models used in the subsequent analyses.
\renewcommand{\arraystretch}{1.3}
\begin{table}
    \centering
    \caption{Parameters common to all simulations.  The minimum frequency and frequency resolution were chosen to match the line-of-sight size of the EoR simulation in \citetalias{S19b}.  To match approximately with the HERA instrument, the time cadence of the simulations was chosen to be 11 seconds.  The number of times was chosen such that the number of data points exceeds the number of model parameters.}
    \begin{tabular}{*5l}
        \enfreq & $\freq_{\rm min}$ & $\Delta\freq$ & \ent & $\Delta t$ \\
        \hline
        38 & 158.3 MHz & 237.6 kHz & 34 & 11 s
    \end{tabular}
    \label{tab:simparams}
\end{table}

\subsection{Array Layout}\label{sec:antenna-layout}

\begin{figure}
    \centering
    \includegraphics[width=0.45\textwidth]{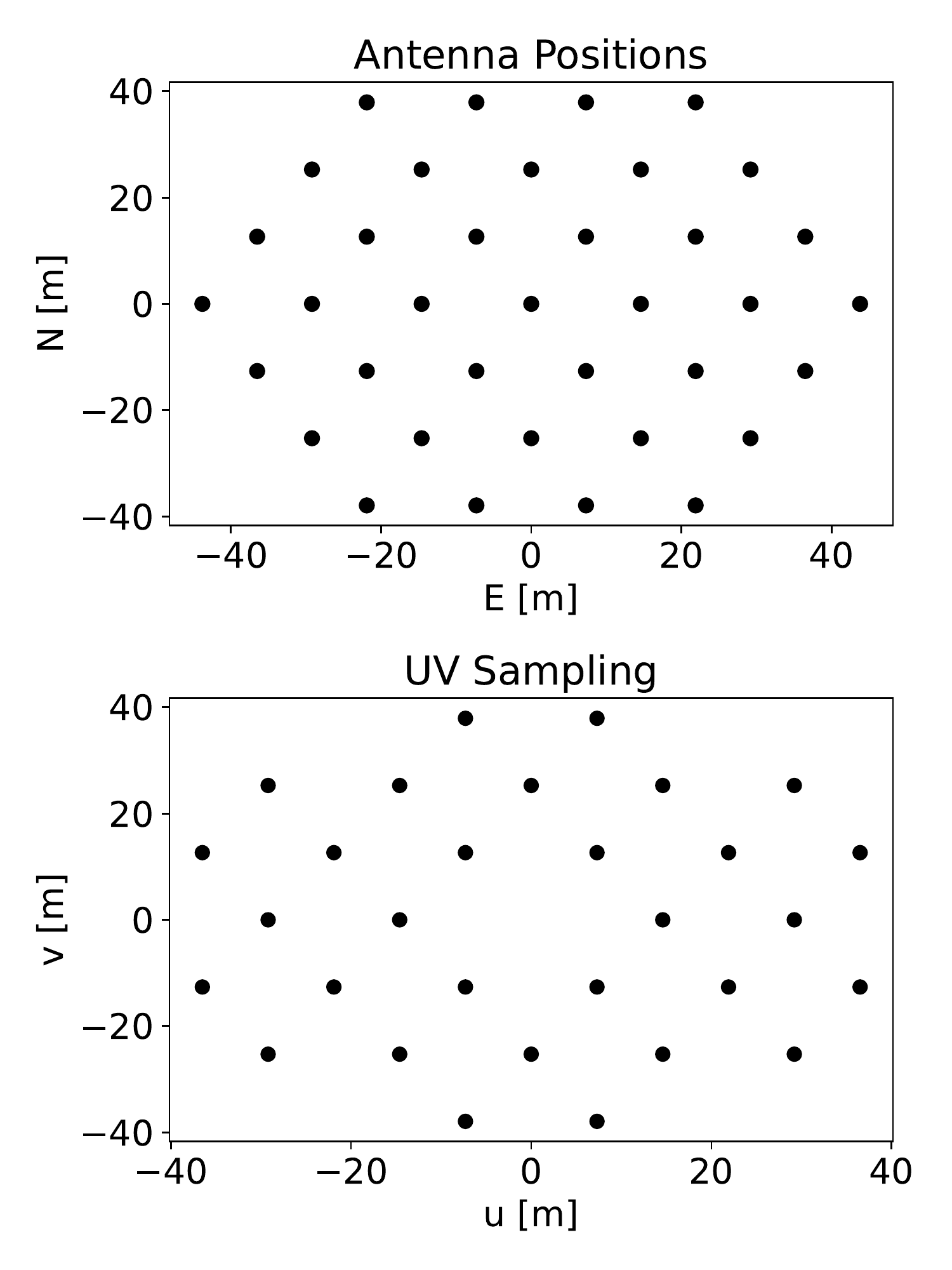}
    \caption{(Top) Antenna positions used in the visibility simulations in (East, North) coordinates defined relative to the array center.  (Bottom) Baselines sampled by the antenna layout in the top panel satisfying $|\bs{b}|\leq40$ m.  The visibilities are modelled as unphased and thus the locations of the visibilities do not change with time.  Baselines at $(u, v)$ are conjugated in the data to $(-u, -v)$ to preserve the Hermitian symmetry in the data vector and the model $uv$-plane.}
    \label{fig:inst-model}
\end{figure}

The array layout was chosen to be a perfectly redundant hexagonal grid with 37 antennas and an antenna spacing and diameter of 14.6 m (see \figref{fig:inst-model}).  Only one baseline from each redundant group with a baseline length $|\bs{b}|\leq 40$ m was simulated.  This baseline cutoff of 40 m was chosen due to computational constraints.  Using longer baselines requires a larger model $uv$-plane which in turn requires longer analysis run times (described in more detail in section \ref{sec:real-data-techniques}).  The center of the array was placed at the location of the HERA array at (lat, lon) $\simeq$ (-30.7$^\circ$, 21.4$^\circ$).

\subsection{Sky Signals}\label{sec:sky-signals}

The sky signals used here fall into two categories: mock EoR and FGs.  Because \beor\ models a subset of the observed sky, two datasets were simulated for each sky signal.  One dataset uses a subset of the sky in the simulation with a FoV that matches the sky model in \beor.  This scenario, where the visibilities only contain contributions from signals located in the patch of sky being modelled, is referred to as the data having a ``restricted FoV''.  The other dataset uses the entire visible sky in the visibility simulation and is thus deemed ``all sky''.  In the restricted FoV scenario, we are assuming perfect knowledge of the FGs outside of the image-space model.  Contrarily, in the all sky scenario, we are assuming no knowledge of the FGs outside of the image-space model.  In these all sky datasets, the visibilities will contain contributions from sources located outside the extent of the sky model.

\subsubsection{EoR}\label{sec:sky-eor}

The mock EoR component was generated as a set of 38 nside=256 \hpx\ maps with amplitudes drawn from a Gaussian distribution $\mathcal{N}(\mu, \sigma)=(0, 6.48$ mK)\footnote{As reionization progresses, the EoR signal becomes increasingly non-Gaussian.  Information about incorporating a non-Gaussian prior into our pipeline to address this can be found in section 3.1 of \citetalias{S19a}.}.  The expected power spectrum amplitude of this mock EoR signal (IID Gaussian noise) is given by
\begin{equation}
    P_{\rm{EoR}}(k) = \sigma^2\,\rmd{V}
\end{equation}
with $\sigma^2$ the variance of the Gaussian distribution and $\rmd{V}$ the voxel volume\footnote{A truly flat power spectrum would require a frequency dependent scaling of $\sigma$ due to the redshift dependence of the voxel volume in Mpc$^3$.  For this work, we are considering a bandwidth of $\sim9$ MHz over which the voxel volume changes by only 6\%, so we ignore this redshift dependence.  The full prescription for obtaining a flat power spectrum for Gaussian noise with the appropriate redshift scaling can be found in \citet{lanman-pober}.}.  For a \hpx\ map, the voxel volume is set by the pixel area $\rmd{A}$ and the frequency channel spacing $\Delta\freq$, i.e. $\rmd{V}=\rmd{A}\Delta\freq$.  The expected dimensionless power spectrum is then
\begin{align}
    \Delta^2_{\rm{EoR}}(k)
    &= \frac{k^3}{2\pi^2}\,P_{\rm{EoR}}(k) \nonumber \\
    &= \frac{k^3}{2\pi^2}\,\sigma^2\,\rmd{V}
    \label{eq:expected-dmps}
\end{align}
and is plotted in all subsequent power spectrum plots as a dashed black line.

\subsubsection{Foregrounds}\label{sec:sky-fgs}

The foregrounds used here are a summation of the 2016 Global Sky Model (GSM, \cite{GSM16}) and the GaLactic and Extragalactic All-Sky MWA Survey (GLEAM, \cite{GLEAM}).
Note that we only included those sources explicitly in the GLEAM catalog, i.e.\ we did not include the brightest point sources which were peeled during the calibration step in forming GLEAM (see Table 2 in \citet{gleam-a-team}).
We used \texttt{PyGSM}\footnote{\url{https://github.com/telegraphic/PyGSM}} to obtain nside=256 HEALPix maps of the GSM at the frequencies in Table \ref{tab:simparams}.
For the GLEAM component, our primary focus was to obtain a point source catalog with representative flux and spatial distributions.  The frequency spectrum of each source, however, was replaced by a power law with a brightness temperature spectral index $\beta$ drawn from a Gaussian distribution $\mathcal{N}(\mu, \sigma)=(2.82, 0.19)$.
The mean and standard deviation of this spectral index distribution matches that used in \citetalias{S19b} obtained from the Very Large Array Low-frequency Sky Survey (VLSS, \citet{beta_egs}).
To set the source spectra, the GLEAM catalog was first interpolated\footnote{Interpolation performed using \texttt{pyradiosky}: \url{https://github.com/RadioAstronomySoftwareGroup/pyradiosky}} using a cubic spline from the native GLEAM spectral resolution ($\sim8$ MHz) to the minimum frequency in Table \ref{tab:simparams}, i.e. 158.3 MHz, to obtain a reference flux value per source.  The spectral structure of each source was then extrapolated using a randomly drawn spectral index.  Each GLEAM source (indexed here by $j$) thus has a flux spectrum in temperature units defined by
\begin{equation}
    I_{j} = I_j^{(0)}\left( \frac{\freq}{\freq_0} \right)^{-\beta_j}
\end{equation}
where $\freq_0$ is the minimum (reference) frequency, $I_j^{(0)}$ is the interpolated GLEAM flux at the minimum frequency, and $\beta_j$ is the randomly drawn spectral index.

The simulated datasets here are centered on (RA, Dec) = (2 hours, -30.7$^\circ$), i.e.\ ``Field 1'' from \cite{hera-uplims}.  This field is centered away from the Galactic plane and was chosen by the HERA collaboration due to the minimal diffuse FGs and the presence of a bright point source for calibration.

\subsection{Beams}

\begin{figure}
    \centering
    \includegraphics[width=0.5\textwidth]{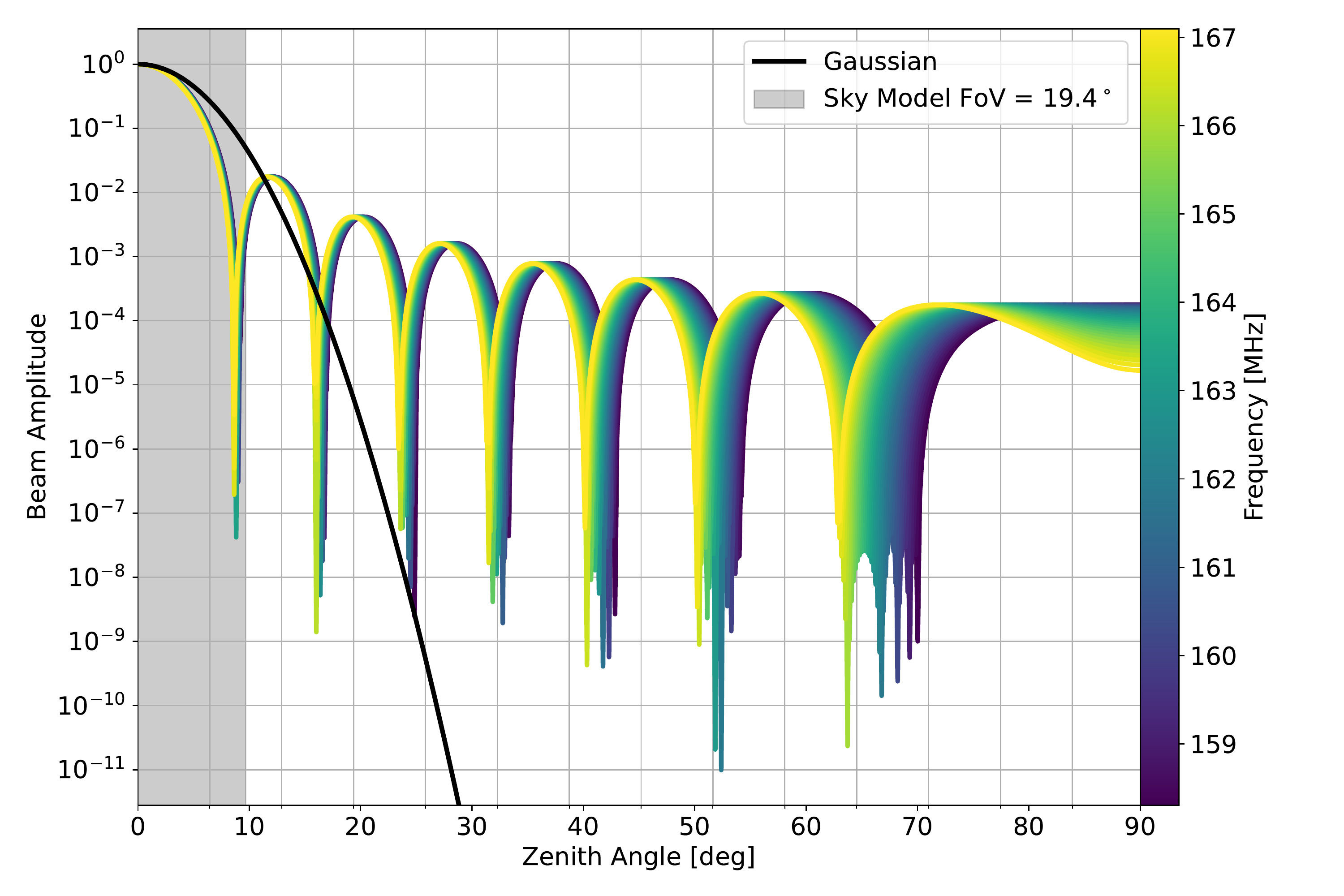}
    \caption{Beam cuts showing the frequency evolution of the Airy beam spatially and spectrally.  The colored lines show a cut through the Airy beam at each frequency in the data from zenith to the horizon.  The solid black line shows a cut through the Gaussian beam for reference.  The grey shaded region shows the extent in zenith angle covered by the image-space model.  The nulls and sidelobes of the Airy beam move spatially with frequency.  As a function of frequency, the nulls far out in the beam change position more drastically than nulls closer to zenith.}
    \label{fig:beam-evolution}
\end{figure}

When marginalizing over the foreground parameters, the sky model in \beor\ is constructed using a limited FoV to improve the computational efficiency.  Visibilities from a realistic instrument, however, contain information about the full sky.  When modelling visibilities using a limited FoV, there are contributions to the visibilities from sources outside the extent of the sky model.  If the beam sufficiently downweights the sources outside of the FoV of the image-space model, then the limited sky model can accurately represent the sky as seen by the instrument.  If the beam has large sidelobes that extend down to the horizon, however, it becomes difficult for the limited sky model to capture the information from the full sky.  To characterize the performance of the limited sky model in the presence of different beam shapes, we simulated visibilities with Gaussian and Airy beams.  A Gaussian beam represents the ideal scenario for a beam with no sidelobe structure.  An Airy beam, on the other hand, represents a more realistic scenario where the beam has significant sidelobe structure.  The Gaussian beam used here was fit to an electromagnetic simulation of the HERA H1C dipole baseline beam \citep{nfbeams} with a fitted FWHM of $\sim$9.3$^\circ$ at 158 MHz.  The Airy beam corresponds to the Airy disk pattern for a circular aperture with diameter 14.6 m.
\figref{fig:beam-evolution} shows a comparison of the two beams used in our analyses as a function of zenith angle.  For either choice of beam, all antennas in a simulation have an identical primary beam response.

\section{Power Spectrum Results}\label{sec:results}

\begin{figure*}
    \centering
    \includegraphics[width=\textwidth]{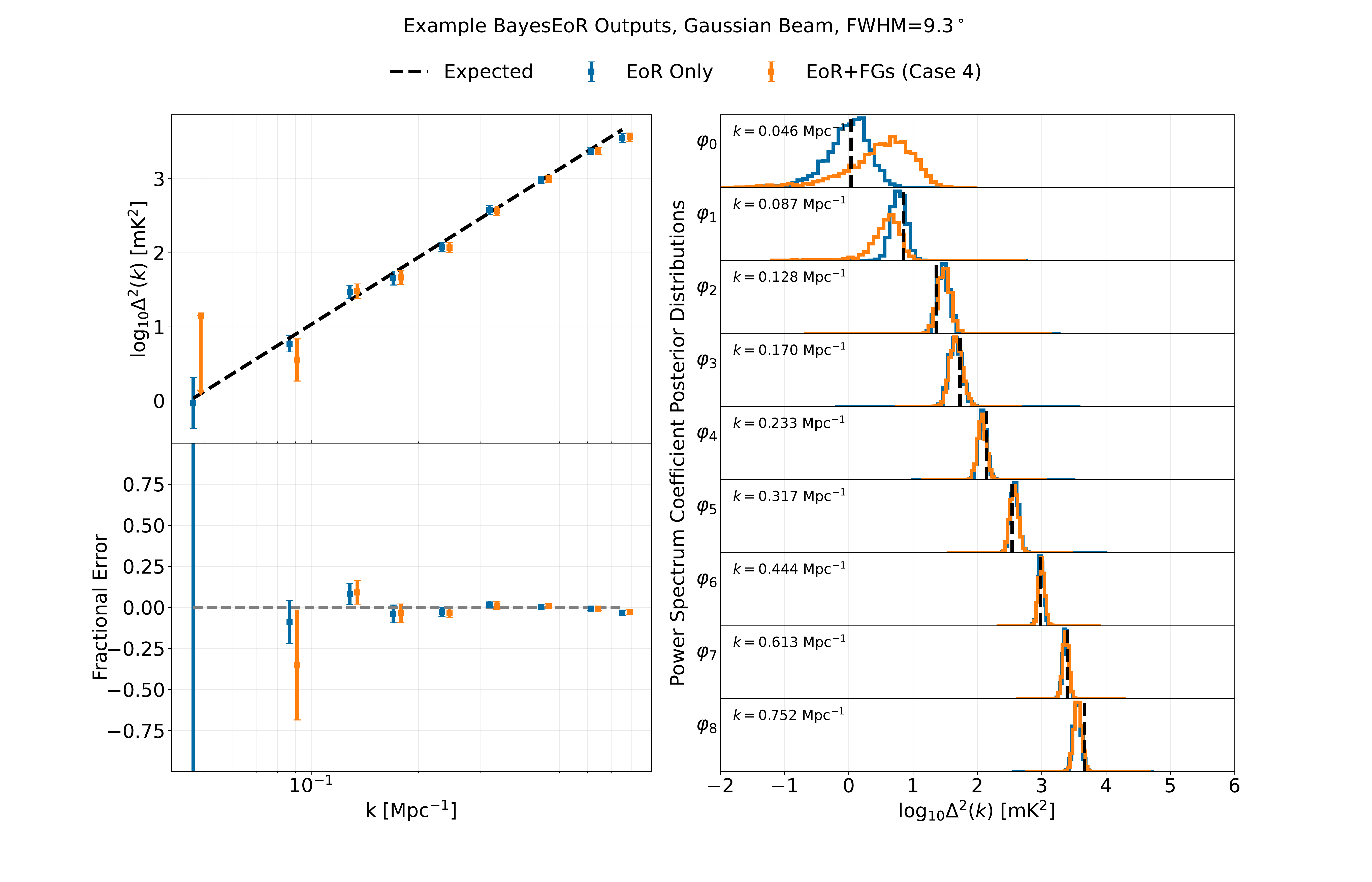}
    \caption{Example plot showing the outputs of \beor.  (Top Left) Recovered posteriors and expected dimensionless power spectrum amplitude plotted as data points with 1$\sigma$ error bars and a dashed black line, respectively.  The data points plotted are the joint-posterior weighted means of each $k$-bin's posterior distribution (seen at right).  The blue and orange data points have been manually offset in $k$ for visual clarity.   (Bottom Left) Fractional difference between data points with error bars and the dashed black line.  (Right) Posteriors for each $k$-bin.  The expected dimensionless power spectrum amplitude is plotted per posterior subplot as a vertical dashed black line.  Plotted in blue is a set of posteriors with detections at all $k$ obtained using log-uniform priors.  Plotted in orange is a set of posteriors with a non-detection obtained with a uniform prior in the lowest $k$-bin and detections using log-uniform priors at all other $k$.}
    \label{fig:summary-plot}
\end{figure*}
A suite of power spectrum tests was conducted using various combinations of the aforementioned sky signals and beams.  These tests fall into three main regimes and are described in the enumerated subsections:
\begin{enumerate}
    \item[\ref{sec:eor-only}] EoR Only
    \item[\ref{sec:eor-fgs-rfov}] EoR+FGs: Restricted FoV
    \item[\ref{sec:eor-fgs-all-sky}] EoR+FGs: All sky
\end{enumerate}
A summary of the results from these tests is available in Table \ref{tab:summary}.  In all subsequent power spectrum plots, unless otherwise noted in the caption, the data points with errorbars are the posterior weighted means of each power spectrum coefficient and the associated standard deviation.  In \figref{fig:summary-plot} we show the detailed results from two example analyses.  These results demonstrate the outputs of \beor: a posterior distribution for each $k$-bin and a joint-posterior probability for each sampled combination of power spectrum amplitudes.
A given $k$-bin is deemed a detection if the difference of the natural log of the evidence for detecting power and no power in that bin is greater than 3.  From \cite{kass-raftery}, this is equivalent to $2\ln(B_{10})>6$, where $B_{10}$ is the Bayes factor comparing the probability of a detection versus a non-detection, which corresponds to ``strong'' evidence (greater than 20:1 odds) in favor of a detection.  For reference, all detections in this work satisfied $\ln(B_{10})>6$ which qualifies as ``very strong'' evidence from \cite{kass-raftery}.
If the difference is less than 3 for a particular $k$-bin, the power spectrum is re-run using a uniform prior on that bin to obtain an upper limit.  All upper limits presented here are calculated as the 95th percentile of the posterior distribution.

\subsection{EoR Only}\label{sec:eor-only}

\begin{figure}
    \centering
    \includegraphics[width=\linewidth]{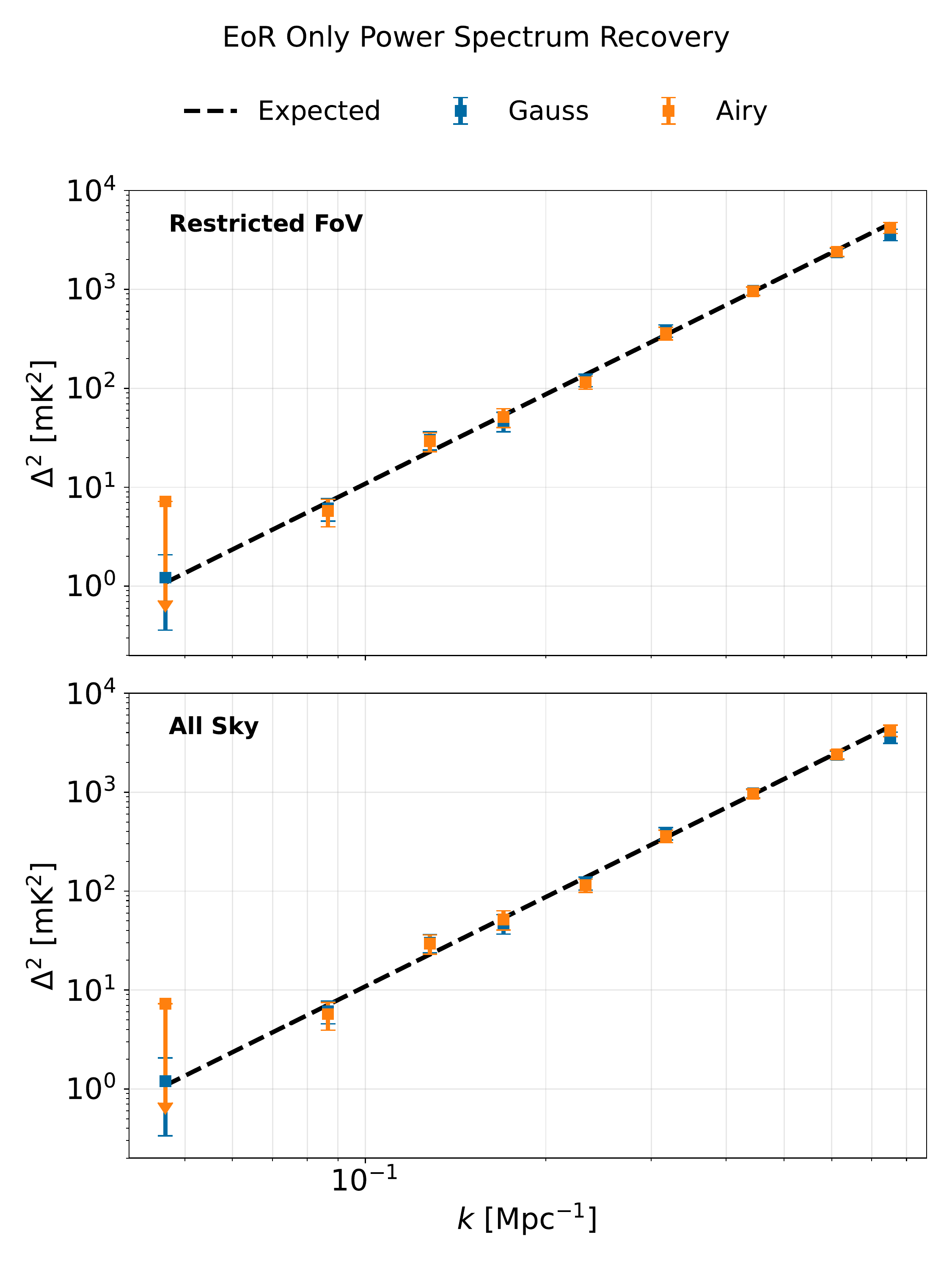}
    \caption{Recovered, spherically averaged dimensionless power spectrum estimates plotted as points with 1$\sigma$ error bars for the EoR only, restricted FoV (top) and all sky (bottom) datasets.  The expected amplitude is plotted as the thick dashed black line as calculated from Equation \ref{eq:expected-dmps}.  The recovered power spectra can be seen to be consistent with the expected power spectrum amplitude at all $k$ for both beam types in both the restricted FoV and all sky scenarios.}
    \label{fig:eor-only}
\end{figure}

Datasets containing only the EoR signal were tested first to ensure that unbiased recovery of the EoR power spectrum is possible without FG contamination.  \figref{fig:eor-only} shows the recovered posterior means and $1\sigma$ errorbars for the EoR only, restricted FoV datasets (top plot).  The recovered power spectra for both beam types agree with the expected power spectrum amplitude in Equation \ref{eq:expected-dmps} at all $k$.  \figref{fig:eor-only} also shows the recovered power spectra for visibilities simulated from an all sky EoR signal (bottom plot).  In the absence of FGs, \beor\ can recover power spectrum estimates consistent with expectation even when modelling all sky visibilities using a subset of the observed sky.

Because our Gaussian and Airy beams downweight the EoR signal outside the model FoV by a factor of $\gtrsim10$ and $\gtrsim100$, respectively, the EoR signal in the all sky data can be well described by the limited image-space model.  However, the extreme ($10^5$) dynamic range between the FGs and EoR signals requires that FGs outside the model FoV be downweighted by a much larger value, i.e.\ $\gg10^2$.  If the beam insufficiently downweights FGs outside the model FoV, this can lead to contamination of the EoR power spectrum estimates (see sections \ref{sec:eor-fgs-all-sky} and \ref{sec:discussion}).

\subsection{EoR+FGs: Restricted FoV}\label{sec:eor-fgs-rfov}

The next set of tests conducted involved EoR+FGs in the restricted FoV scenario.  In this regime, the information content of the sky should be completely describable by the image-space model.  The results of these tests can be seen in Table \ref{tab:summary} as cases 1 and 2 for the Gaussian and Airy beam, respectively.  The recovered power spectra for both beam types were consistent with the expected power spectrum amplitude at all $k$.  These results demonstrate that if the FG model is sufficient to describe the FGs in the data, unbiased EoR power spectrum estimates can be recovered at all $k$.  This is true even when using an Airy beam containing spatial and spectral structure within the image-space model.

\subsection{EoR+FGs: All Sky}\label{sec:eor-fgs-all-sky}

When using a Gaussian beam (Case 3 in Table \ref{tab:summary}), the all sky EoR+FG dataset once again returned power spectrum estimates consistent with the expected power spectrum at all $k$.  The Airy beam (Case 4), however, did not.  The contamination for the Airy beam all sky EoR+FG dataset was such that the priors had to be expanded to a range where the analysis became numerically unstable.  The failure of this particular combination of Airy beam and all sky FGs is due to insufficient downweighting of bright FGs far from zenith (described below in section \ref{sec:discussion}).

\section{High Zenith Angle FG Modelling Errors}\label{sec:discussion}

By construction, the image-space model is restricted in its FoV and can only model a subset of the delays and fringe rates present in the all sky visibilities.  Sources far from zenith will thus have delays and fringe rates that are inaccessible via the image-space model.  The beam must downweight these bright, high zenith angle sources sufficiently for the image-space model to adequately describe the all sky visibilities.

\begin{figure}
    \centering
    \includegraphics[width=0.5\textwidth]{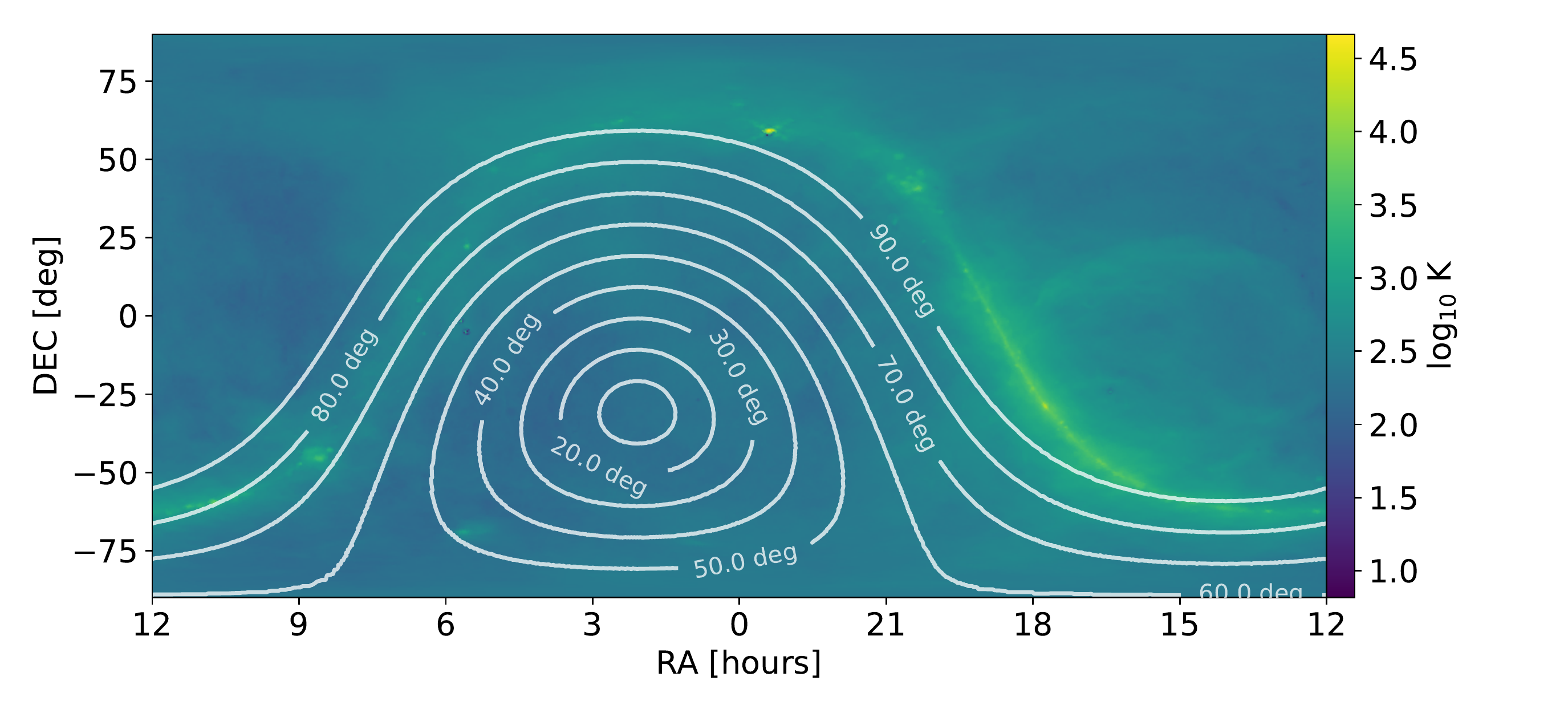}
    \caption{GSM overlaid with zenith angle contours from the central time step in the simulated visibilities.  The LMC is visible at (RA, Dec) $\simeq$ (5.5 hours, -70$^\circ$) in between the 40 and 50$^\circ$ zenith angle contours.  M42 is visible at (RA, Dec) $\simeq$ (6 hours, 0$^\circ$) in between the 50 and 60$^\circ$ zenith angle contours. The galactic anti-center lies above the horizon at $\gtrsim70^\circ$ in zenith angle.}
    \label{fig:gsm-contour}
\end{figure}

\begin{figure}
    \centering
    \includegraphics[width=0.5\textwidth]{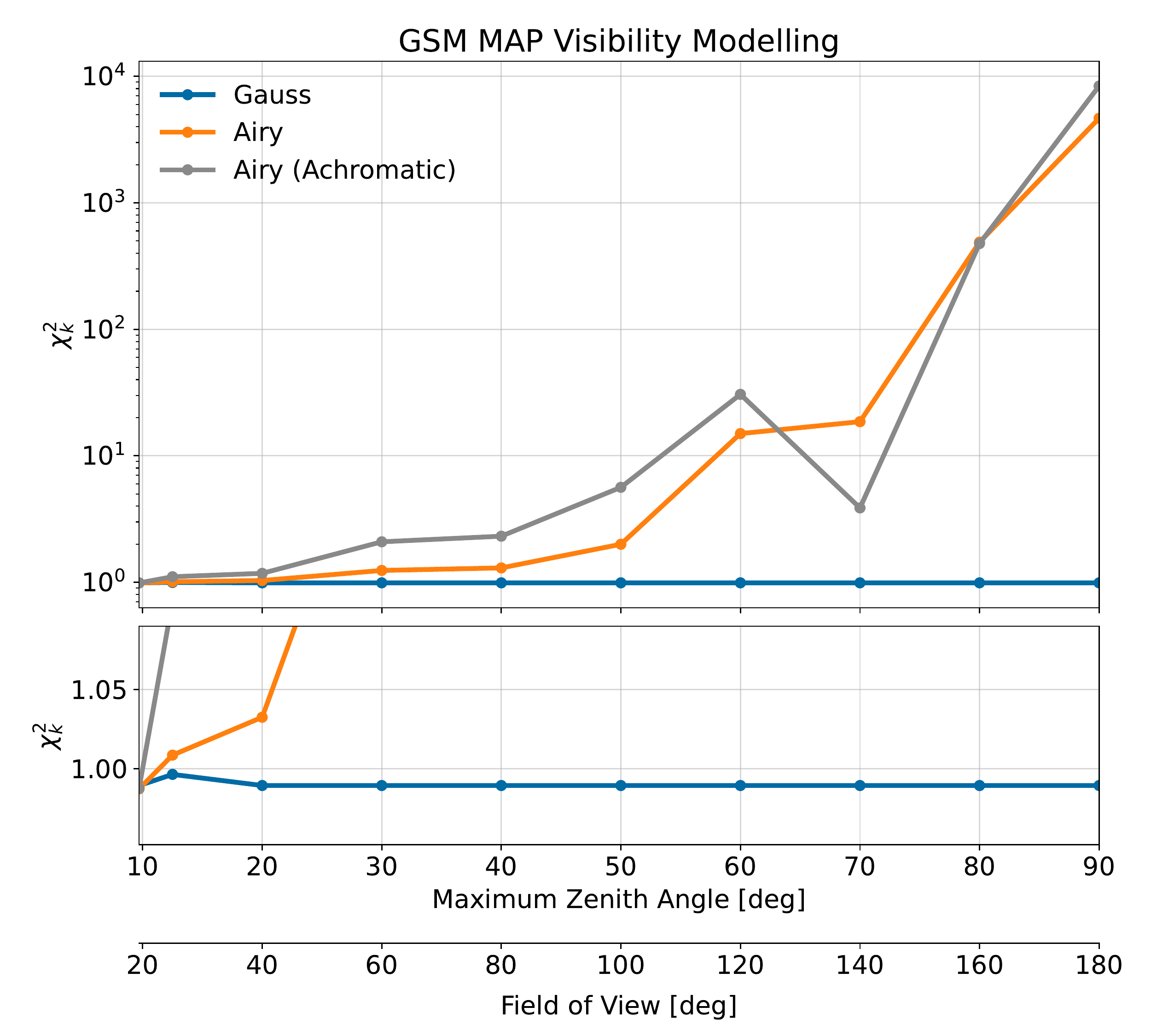}
    \caption{(Top) Reduced $\chi^2$ values (computed via Equation \ref{eq:chisq}) versus the FoV of the input GSM only visibilities.  (Bottom) Zoom in of the top plot to illustrate the small deviations in the Airy beam results (orange and grey lines) from the Gaussian result (blue line).  The upper and lower $x$-axes detail the maximum zenith angle in and the associated FoV of the GSM only visibilities.  The maximum zenith angle has been included for convenience to compare with the zenith angle contours in \figref{fig:gsm-contour}.  For the Airy beams (both chromatic in orange and achromatic in grey), the modelling error increases as larger FoV values are used in the simulated visibilities.  This indicates that the limited image-space model is insufficient to describe the visibilities.  For the Gaussian beam, the modelling error is roughly constant.  The Gaussian beam downweights the sky drastically compared to the Airy beam.  The sources far out in the beam which are difficult to model thus contribute less to the Gaussian beam visibilities.}
    \label{fig:map-gsm-model-errors}
\end{figure}

To confirm this, we simulated a suite of GSM only visibilities, $\mathbf{V}_{\rm{GSM}}$, using a range of FoV values from 19.4$^\circ$ (restricted FoV) to 180$^\circ$ (all sky).  The FoV in each dataset sets the maximum zenith angle of the GSM included in the visibility simulation.  For example, a dataset with a FoV of 20$^\circ$ only includes contributions from the GSM out to 10$^\circ$ in zenith angle (see \figref{fig:gsm-contour}).  Simulations with a FoV larger than that of the image-space model (fixed at 19.4$^\circ$) thus contain information about the sky outside the extent of the model.  For each set of input visibilities (each FoV value), we computed a set of maximum a posteriori (MAP) visibilities using only the FG model basis vectors, $\mathbf{V}_{\rm{MAP}}$.
\figref{fig:map-gsm-model-errors} shows the performance of the MAP fit as a function of the FoV of the input visibilities. We use the reduced $\chi^2$ statistic computed as
\begin{equation}
    \chi_k^2 =
        \frac{1}{k}
        \sum_{i=1}^{k} \frac{\left| (\mathbf{V}_{\rm{GSM}, i} + \nvec_i) - \mathbf{V}_{\rm{MAP},i} \right|^2}{\sigma_{\nvec}^2}
    \label{eq:chisq}
\end{equation}
to assess the goodness of fit of the MAP visibilities.  In the equation above, the subscript $i$ indexes the $i$-th visibility in the visibility vector $\mathbf{V}_{\rm{GSM}}$, for example, with $k=N_{\rm{vis}}\sim38,000$ visibilities (degrees of freedom).  If the MAP fit to the GSM only visibilities is good, then the residuals $(\mathbf{V}_{\rm{GSM}} + \nvec) - \mathbf{V}_{\rm{MAP}}$ should produce a distribution consistent with the noise, i.e. normally distributed with mean zero and variance $\sigma_{\mathbf{n}}^2$.  In this case, $\chi_k^2\sim1$.  Because the number of degrees of freedom (visibilities) is so large, small changes in $\chi_k^2$ result in large modelling errors.  For the Airy beam (orange line in \figref{fig:map-gsm-model-errors}), $\chi_k^2$ increases monotonically with the FoV.  Relative to the minimum FoV (19.4$^\circ$), the small deviations in $\chi_k^2$ at modest FoV values of 25$^\circ$ and 40$^\circ$ seen for the Airy beam represent significant FG modelling errors that result in EoR power spectrum contamination (see the bottom plot in \figref{fig:map-gsm-model-errors}).  Notable increases in $\chi_k^2$ can be seen at higher zenith angle values of 50$^\circ$, 60$^\circ$ and 80$^\circ$.  As seen in \figref{fig:gsm-contour}, these zenith angle values coincide with the locations of bright FG sources like the Large Magellanic Cloud (LMC, zenith angle $\sim50^\circ$), M42 (zenith angle $\sim55^\circ$) and the galactic plane (zenith angle $\sim75^\circ$).  These bright FG structures lie well outside the image-space model.  The modelling error for the Gaussian beam, on the other hand, is approximately fixed for all values of the FoV.  This is precisely because the Gaussian beam strongly downweights the sky outside the extent of the image domain model.  For comparison, \figref{fig:beam-evolution} shows the amplitudes of the Gaussian and Airy beams as a function of zenith angle.  The minimum amplitude of the Airy beam (ignoring the nulls) is obtained at the horizon at a level of $\sim10^{-4}$.  The Gaussian beam reaches this same amplitude at a zenith angle of only $\sim17^\circ$.  The contributions from bright FGs far out in the Gaussian beam are thus negligible.  The Airy beam, however, does not sufficiently downweight these bright sources which results in a poor fit of the all sky visibilities using the limited image-space model.
In constructing a set of model visibilities, the code does what it can to most accurately reproduce the input visibilities.  When modelling all sky visibilities with a beam sensitive to high zenith angle FG emission, however, there is a significant contribution to the input visibilities from sources outside the extent of the image domain model.  Ignoring this unmodelled FG power completely produces a much worse fit than allowing that power to leak into the low $k$ modes used to model the EoR signal.  In this way, unmodelled FG emission from outside the extent of the sky model ends up in the Fourier modes used to estimate the EoR power spectrum and contaminates the power spectrum estimates.

\figref{fig:beam-evolution} also shows the spectral evolution of the Airy beam as a function of zenith angle.  As a function of frequency, the nulls far out in the beam change position more drastically than nulls closer to zenith.  A source at a fixed zenith angle far out in the beam will thus have additional spectral structure imparted on it by the spectral evolution of the beam.
Because the image-space model only models the beam out to zenith angles of $\sim9.7^\circ$, FGs far out in the beam could have spectral structure not well described by the FG model.
Specifically, if the beam imparts spectral structure onto the FG signal on scales used to model the EoR, the FG power will be scattered by the beam to higher frequency Fourier modes, i.e.\ higher $k_z$.  Because we do not model the beam at these high zenith angles, this results in beam modulated FG power in the data being absorbed by Fourier modes used to model the EoR signal which contaminates the EoR power spectrum estimate.
To disentangle the effects of the beam chromaticity, an identical set of tests to those described in the previous paragraph was conducted using an achromatic version of the Airy beam.  The achromatic Airy beam was formed by fixing the spatial structure of the beam at the lowest frequency in the data.  In this case, the results were consistent with the chromatic beam (see the grey line in  \figref{fig:map-gsm-model-errors}).  This implies that the chromaticity of the beam is not the dominant source of error when modelling the all sky visibilities with an Airy beam.

\renewcommand{\arraystretch}{1.1}
\begin{table*}
\caption{
    Table summarizing $k$-modes that can be recovered in an unbiased manner with different modelling approximations and observing assumptions.  Each analysis scenario is assigned a case number (left column).  The remaining columns from left to right describe: whether the sky signal was ``restricted`` to the FoV of the model, an abbreviation of the beam type used (\textbf{G}aussian, \textbf{A}iry, \textbf{T}apered \textbf{A}iry), whether or not any FG mitigation strategy was applied prior to power spectrum analysis, and whether each $k$-bin in the analysis reported was an upper limit (down arrow), detection consistent with expectation (check mark), or contaminated (no symbol).  The numerical value for each $k$-bin referenced here can be found in \figref{fig:eor-only}.
    \vspace{0.1in}
}
\centerline{
    \begin{tabular}{l l l l l l l l l l l l l }
        Case & Restricted & Beam & FG & \multicolumn{9}{c}{$k$-bin} \\
        & FoV & & Mitigation & 0 & 1 & 2 & 3 & 4 & 5 & 6 & 7 & 8 \\[0.1cm]
        \hline
        & & & & & & & & & & & & \\[-0.3cm]
        1 & Y     & G & N
        & $\downarrow$ & \checkmark & \checkmark & \checkmark & \checkmark & \checkmark & \checkmark & \checkmark & \checkmark \\
        2 & Y     & A & N
        & $\downarrow$ & $\downarrow$ & \checkmark & \checkmark & \checkmark & \checkmark & \checkmark & \checkmark & \checkmark \\
        3 & N     & G & N
        & $\downarrow$ & \checkmark & \checkmark & \checkmark & \checkmark & \checkmark & \checkmark & \checkmark & \checkmark \\
        4 & N    & A & N
        &  &  &  &  &  &  &  &  &  \\
        5 & N     & TA & N
        & $\downarrow$ & \checkmark & \checkmark & \checkmark & \checkmark & \checkmark & \checkmark & \checkmark & \checkmark  \\
        6 & N     & A & Y
        & $\downarrow$ & \checkmark & \checkmark & \checkmark & \checkmark & \checkmark & \checkmark & \checkmark & \checkmark  \\
    \end{tabular}
}
\label{tab:summary}
\end{table*}

\section{Techniques for Real Data}\label{sec:real-data-techniques}

Modelling visibilities from an instrument that can see the entire sky via a sky model encompassing a subset of the sky is difficult.  As demonstrated above, an image-space model with a FoV < 180$^\circ$ is susceptible to modelling errors when using data from an instrument with a broad beam.  This result isn't exactly surprising, however.  Previous works have demonstrated that FG mitigation is a truly wide-field problem and both our instruments and analyses must be sophisticated enough to address these wide-field effects \citep{pober, thyagarajan}.  While using a larger model FoV is possible, it comes at the cost of increased computational expense due to large matrix inversions (discussed in section \ref{sec:increase-model-fov}).  This large matrix inversion can be avoided by exploring the full un-marginalized parameter space, but this approach comes with its own trade-offs.

While analytically marginalizing over the signal coefficients \avec\ and \qvec\ allows us to explore a much smaller parameter space, it is not strictly required.  Using the joint posterior prior to marginalization avoids the need for the large matrix inversion step.  It does, however, require sampling over the much larger parameter space of all {\avec}s and {\qvec}s.  This in turn requires sampling techniques designed to efficiently explore large parameter spaces.  \citetalias{S19a} provides a demonstration of \beor\ using the un-marginalized joint-posterior and a Guided Hamiltonian Sampler (GHS, see sources within section 5 of \citet{S19a}).  However, marginalizing over the FG parameters has the significant advantage of greatly reducing the size of the parameter space.  As mentioned in the introduction, with a smaller parameter space we can employ nested sampling which facilitates the use of powerful Bayesian model selection tools unavailable to other sampling techniques (see e.g.\ \citetalias{S19b}).  Thus, here, we discuss updates to the data model that allow us to retain these advantages while keeping the analysis computationally tractable.  Below, we propose several ways in which our analysis could be updated to recover unbiased power spectral estimates when marginalizing over the FG parameters such as increasing the complexity of the FG model, modifying the likelihood to compare data and model in the image domain, mitigating FGs in the data prior to power spectrum estimation, and designing instruments with narrow beams.

\subsection{Increase FG Model Complexity}\label{sec:increase-model-fov}

Ideally, the image-space model would be able to encompass the entire visible sky.  Expanding the FoV of both the EoR and FG components of the model to 180$^\circ$ is infeasible, though, due to memory requirements.  As described in section \ref{sec:model-params}, for a fixed instrument model, increasing the sky model FoV requires a larger value of \enu\ ($\rmd{u}\propto\rm{FoV}^{-1}$).  This in turn requires a larger matrix inversion per posterior calculation which scales as \enu$^3$.  For reference, with $\enu=23$ and FoV$\ =19.4^\circ$, a single matrix inversion takes $\sim$10 seconds on a Tesla p100 GPU.  For the parameter space explored here, this would result in an average run time for a single power spectrum analysis of approximately 15 hours when run in parallel on 3 GPUs (45 GPU hours).  Modelling the entire sky (FoV$\ =180^\circ$) would require \enu$=213$ to encompass the baselines simulated in this work.  At this scale, a single matrix inversion would take $\gtrsim$2 hours and the total run time on the 6 GPUs available to us would be approximately 250 days (1500 GPU hours).  On a large computing cluster with more GPUs, a single all sky analysis could be run on a week to month timescale.  Number of GPUs aside, holding the all sky matrix in memory is unrealistic, requiring $\sim52$ TB of RAM.  Even on an ACCESS Bridges-2 class supercomputer\footnote{\url{https://www.psc.edu/resources/bridges-2/}}, this memory requirement is prohibitively large.  Without modification, expanding the FoV of both the EoR and FG components of the model is thus computationally infeasible.

Figure \ref{fig:eor-only} shows that the limited image-space model is sufficient to recover the EoR power spectrum even when modelling an all sky EoR signal.  When analyzing all sky EoR+FG data, however, the FG model requires a larger FoV than used here (19.4$^\circ$) to properly model bright FGs far from zenith.  A natural way to improve the FG modelling capabilities without sacrificing computational efficiency is to increase the FoV of the FG model while keeping the FoV of the EoR model fixed.

The current FG model consists of the $\eta=k_z=0$ slice of the $k$-cube plus two power law coefficients in frequency for each $(u, v)$ in the model $uv$-plane (CPDPL model, see Equation \ref{eq:cpdpl}), and the monopole $(u, v) = (0, 0)$ per $k_z$.  For the sake of this discussion, we will ignore the $(u, v) = (0, 0)$ component of the FG model as it is unaffected by the following changes.  Currently, increasing the FoV of the image-space model requires increasing \enu\ for both the EoR and FG models simultaneously which increases the runtime as \enu$^3$.  Alternatively, it is possible to define a separate set of parameters ($N_u^{\rm{FG}}$, FoV$_{\rm{FG}}$) for the FG model only.  In this way, only the size of the $k_z=0$ slice would be modified, and the remainder of the $k$-cube (used to model the EoR) would remain unchanged.  The FG model could thus utilize a larger FoV without incurring the added computational expense of expanding the FoV of the whole model.  With this modification, the all sky FG model with $N_u^{\rm{FG}}=213$ would require a $\sim400$ GB matrix with shape $(\enmodel, \enmodel)$.  The number of model parameters (\enmodel) is determined by the parameters \enu, \env, \eneta, and \enq.  The total number of model coefficients is determined by the sum of the sampled EoR and FG model modes, (\enu\,\env - 1$)\cdot($\eneta - 1) and $N_u^{\rm{FG}}\,N_v^{\rm{FG}}\cdot($1 + \enq) + \eneta - 1, respectively.  While this memory requirement is too large given our current resources, it is easily within memory limits for large scale computing clusters.  The Matrix Algebra on GPU and Multicore Architectures (MAGMA, \citet{magma}) library employed in \beor\ is already capable of partial i/o.  The only limiting factor for an all sky FG model analysis then is the need for more memory.

Alternatively, if the majority of the unmodelled FG power stems from a few bright sources, it is in principle possible to model a limited number of patches of sky outside the extent of the EoR image domain model.
This approach would not require expanding the FoV of the FG model to span the entire sky or changing the number of model parameters.
The efficacy of this approach, however, is dependent upon two factors: the dominant source of FG emission (diffuse or point source) and the residual FG signal amplitude after fitting for the flux of a limited set of bright sources.  What FG signal dominates depends upon the baseline lengths of interest.  All baseline lengths are sensitive to point source emission, but diffuse emission dominates on short baselines.  For a fixed FoV of the EoR image domain model, modelling longer baselines requires larger \enu\ as $\Delta u = {\rm FoV}^{-1}$.  We are thus restricted to modelling shorter baselines where diffuse emission dominates to ensure reasonable analysis run times.  Assuming we removed the FG flux from a set of bright point sources and bright localized sources like the LMC and M42, there will still be residual unmodelled diffuse FG flux in the visibilities that the model cannot describe.  As we will see in section \ref{sec:fg-mitigation-egsm}, even at a reduced amplitude, this residual diffuse signal leads to corruption of the EoR power spectrum.  Additionally, as seen in \figref{fig:map-gsm-model-errors}, even diffuse emission just outside of the model FoV causes FG modelling difficulties.  Furthermore, the plane of the galaxy extends over a large region of the sky.  Because the galactic plane causes large FG modelling errors (see the modelling error at high zenith angles in \figref{fig:map-gsm-model-errors}), it might be necessary to model large patches of the sky.  If the number of patches and pixels approaches the all sky FG modelling scenario, then this approach is no longer viable.  Ultimately, the efficacy and efficiency of this alternate approach requires further investigation and is left as future work.

\subsection{FG Mitigation}\label{sec:fg-mitigation-egsm}

Instead of increasing the complexity of the FG model, an alternative approach to improve power spectrum estimates involves subtracting a model of the FGs from the visibilities and fitting the power spectrum of the residuals.
This would mitigate a majority of the FG flux in the data which would in turn increase the quality of the FG fit.
LoFAR and the MWA also remove FG emission prior to 21 cm power spectrum estimation (see \citet{lofar-limits} and \citet{mwa-limits} and references within).  However, an important distinction between those approaches and our approach here is that we are not subtracting a FG model derived from the data.  Instead, we are subtracting a fixed FG model derived from a priori knowledge of the instrument and sky, thus negating the risk of signal loss associated with subtracting a fitted sky model.

To test this technique, we used an nside=128 \hpx\ map of estimated 1$\sigma$ errorbars from the extended Global Sky Model (eGSM, \citeauthor{egsm} (in prep.)).  In using this eGSM error map to simulate visibilities, we are effectively assuming that we can subtract off a FG model from the data that produces uncertainties on our FGs consistent with eGSM error bars.  Because the eGSM error map should contain point source emission below the confusion limit of the map, we are also assuming that these errors contain information about unmodelled point sources.  Thus, this error map should encapsulate uncertainties on both diffuse and point source emission.

\begin{figure}
    \centering
    \includegraphics[width=0.5\textwidth]{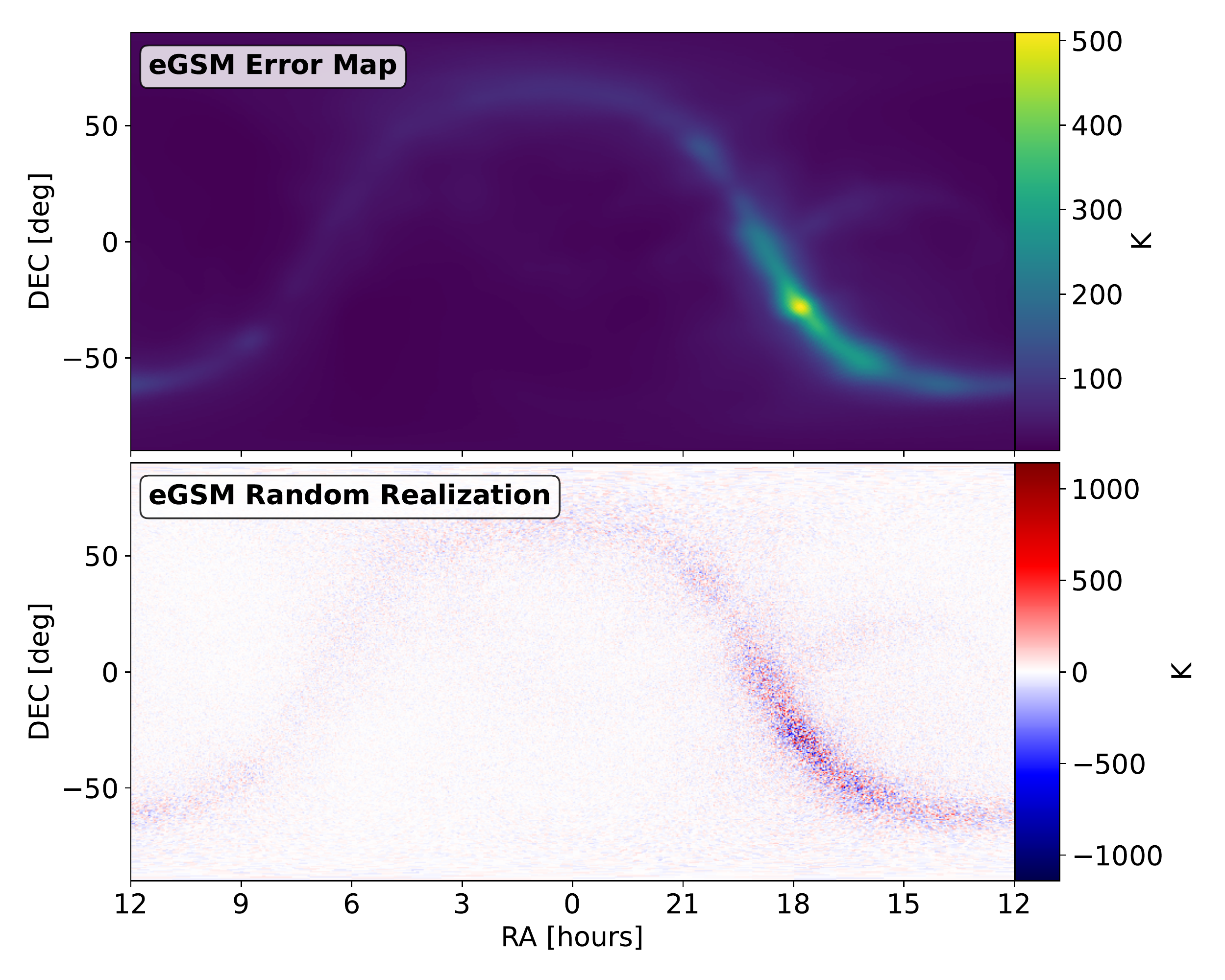}
    \caption{(Top) eGSM Error Map $1\sigma$ error bar estimates.  (Bottom) Random realization drawn using the eGSM Error Map.  This random realization was obtained by drawing a per-pixel sample from a zero mean Gaussian distribution with a standard deviation equal to the amplitude of the corresponding error bar in the top panel.  Both panels display the map at the minimum frequency in the simulation (158.3 MHz).}
    \label{fig:egsm-error-map}
\end{figure}

The actual \hpx\ map that went into the simulation of the eGSM error map visibilities was drawn as a random realization from the original error map.  This randomized realization is a complimentary nside=128 \hpx\ map with a per-pixel amplitude drawn from a Gaussian distribution with mean zero.  The standard deviation of the distribution for each pixel was set as the corresponding pixel amplitude in the error map.  The spectrum of this realization was extrapolated using a power law spectrum with a per-pixel spectral index drawn from a distribution $\mathcal{N}(\mu, \sigma)=\mathcal{N}(2.725, 0.03)$.  The mean spectral index was chosen as the mean of the two spectral indices in the LSSM.  The standard deviation was chosen such that a majority of the spectral indices lie within $\left<\beta\right>_{\rm{GDSE}}$ and $\left<\beta\right>_{\rm{EGS}}$.  The original eGSM error map and the realization used in the visibility simulation can be seen in \figref{fig:egsm-error-map}.

It is important to note that the realization of the eGSM errors in the bottom panel of \figref{fig:egsm-error-map} lacks the proper spatial correlation.  Getting an accurate map of correlated errors requires a full covariance matrix which was unavailable to us.  It is difficult to say whether excluding the spatial correlation from our random realization of the eGSM error map represents an optimistic or pessimistic scenario.  This is due to the complex interplay between the baseline fringe pattern and source positions which is a time, frequency, and baseline dependent effect.  Where FGs land within the fringe pattern ultimately determines whether the adjacent FG pixel values add coherently or incoherently with the fringe pattern.  We leave a more in-depth investigation of the impact of correlated errors to future work.

As opposed to the case of all sky Airy + EoR + GSM \& GLEAM (case 4 in Table \ref{tab:summary}), using an Airy beam with all sky EoR \& eGSM errors resulted in recovered power spectrum estimates at all but the lowest k-bin.  This is a marked improvement over case 4 which possessed FG contamination in the power spectrum so large the analysis became numerically unstable.  The power spectrum recovery is still imperfect, however, producing a FG contaminated power spectrum estimate at low $k$.  While the reduction in FG amplitude from using the error map as opposed to GSM+GLEAM greatly reduces the contamination in the power spectrum, the issue still stands that bright FG sources at high zenith angles are insufficiently downweighted by the beam.  The presence of the galactic plane far outside the of the image-space model still poses modelling problems, even at this reduced amplitude.

\begin{figure}
    \centering
    \includegraphics[width=0.5\textwidth]{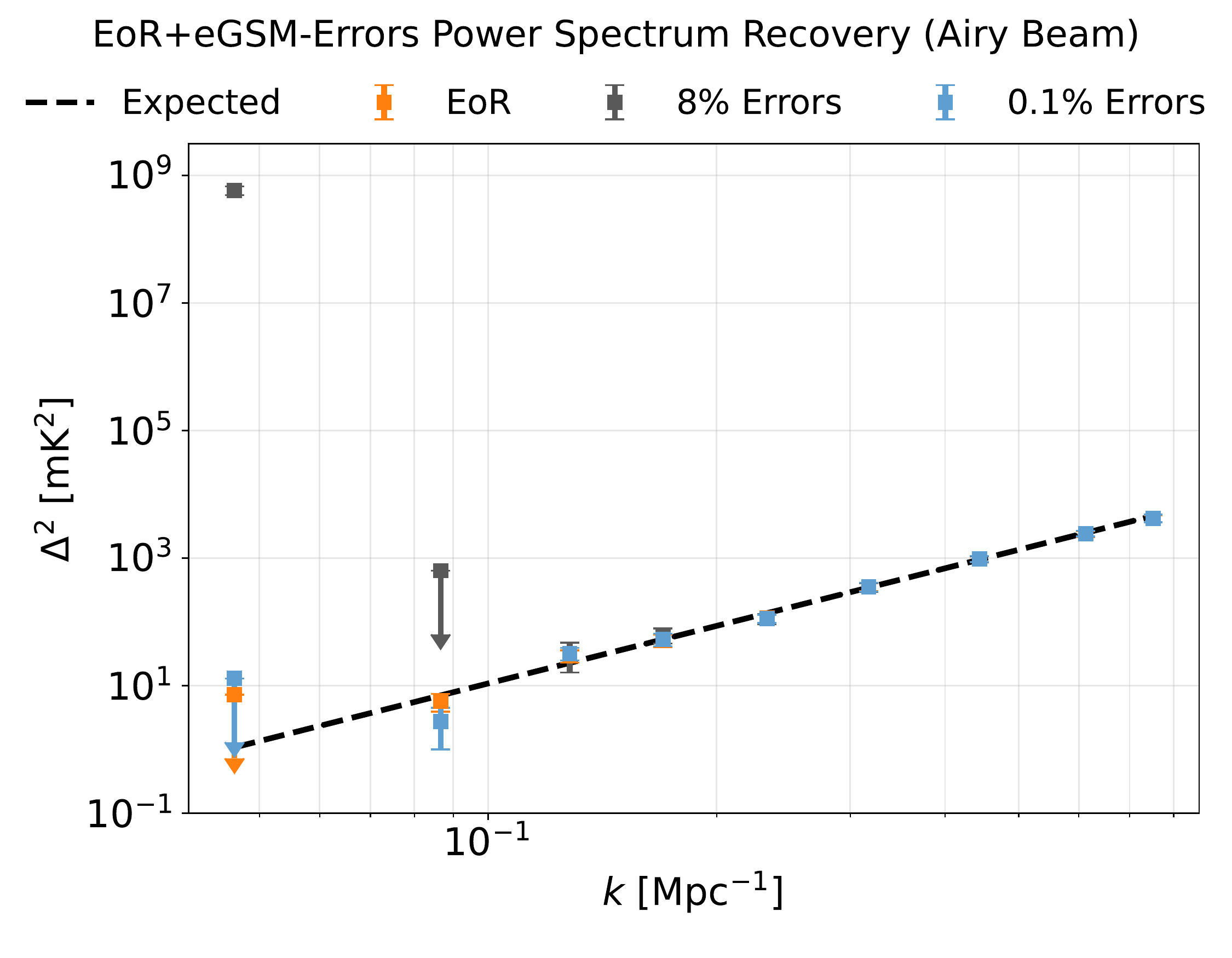}
    \caption{Recovered, spherically averaged dimensionless power spectrum estimates and 1$\sigma$ error bars for all sky Airy beam + eGSM-Error-Map visibilities.  The recovered EoR only, Airy beam power spectrum from the bottom panel of \figref{fig:eor-only} is plotted here in orange for reference.  The lowest $k$-bin is severely contaminated by unmodelled FG emission when using fractional errors of 8\% (dark grey data points), i.e. the error amplitude seed in \figref{fig:egsm-error-map}.  When reducing the fractional errors to 0.1\% (light blue data points), the power spectrum estimates are consistent with expectation at all $k$.}
    \label{fig:egsm-error-ps}
\end{figure}

Relative to the 2016 GSM used here, the errors in the top panel of \figref{fig:egsm-error-map} represent $\sim8\%$ fractional errors.  To recover a power spectrum consistent with expectation at all $k$, we had to reduce the eGSM error bars by a factor of 80 ($0.1\%$ fractional errors, case 6 in Table \ref{tab:summary}).  This implies we would need to know our FGs at the $\sim0.1\%$ level to obtain unbiased EoR power spectrum estimates when using a realistic (Airy) beam and a sky model FoV of 19.4$^\circ$.  While this level of FG knowledge is unrealistic given our current error bars (\figref{fig:egsm-error-map}), we provide this result to demonstrate that improved FG models could allow us to use a narrower model FoV which reduces our computational expense.

\subsection{Image Domain Likelihood}
\label{sec:image-domain-likelihood}

Instead of comparing the data and model in measurement-space, we could in principle compare the two in image-space.  Techniques like those presented in \citet{dom} are being developed for direct optimal mapping of 21 cm data.  This approach offers a way of obtaining an image from visibilities while preserving the power spectrum.  Comparing in the image domain would allow us to multiply both image and model by a tapering function to downweight high zenith angle emission (similar to the approaches in \citet{TGE} and \citet{barry}).  This would potentially permit smaller values for the model FoV (compared to modelling the whole sky) saving drastically on computational expense.  As it stands, the log likelihood in \beor\ (see Equation \ref{eq:likelihood} for reference) is of the form
\begin{equation}
    \log\mathcal{L}_{vis}
    \propto
    (\dvec - \mathbf{T}\bs{\theta})^T\Ninv(\dvec - \mathbf{T}\bs{\theta})
\end{equation}
From \citet{dom}, the direct optimal mapping from visibilities to tapered image can be described as a matrix product $\mathbf{D}\mathbf{A}^\dagger\Ninv$.  Here, $\mathbf{D}$ is a diagonal matrix describing the relative weight of each pixel in the image domain, $\mathbf{A}$ is the measurement matrix relating the \hpx\ gridded sky to the beam convolved visibilities (akin to our \Finvm), and $\mathbf{N}$ is the covariance matrix of the visibilities. Using this matrix product, which for brevity we will call $\mathbf{X}$, the log likelihood comparing data and model in the image domain would take the form
\begin{equation}
    \log\mathcal{L}_{im}
    \propto
    (\mathbf{X}\dvec - \mathbf{X}\mathbf{T}\bs{\theta})^T\mathbf{X}\Ninv\mathbf{X}^T(\mathbf{X}\dvec - \mathbf{X}\mathbf{T}\bs{\theta})
\end{equation}
Here, $\mathbf{X}\Ninv\mathbf{X}^T$ is the covariance matrix of the image-space representation of the data.

A potential caveat with this approach is that the resulting model and data images are equal to the convolution of the intrinsic sky with the point spread function (PSF) of the array.  Ideally, we would combine this approach with the FG mitigation approach from the previous subsection, i.e.\ use input visibilities which have had a fixed FG model (e.g.\ the eGSM in \figref{fig:egsm-error-map}) subtracted prior to analysis with \beor.  Differences in the true and fixed model skies outside of the model FoV will therefore bias the image domain fit within the model FoV.
The bias will thus be proportional to (i) the difference between the true sky and the fixed model sky outside the modelled FoV and (ii) the amplitude and extent of the PSF.  The sparser the $uv$-coverage, the more dispersed the PSF \citep{razavi-ghods, tms}.  Correspondingly, sources further from the modelled FoV contribute more significantly to flux in the model FoV when using this approach with a redundant array such as HERA.

Additionally, when analyzing real datasets, we will need to fit for the noise amplitude to avoid biasing the estimated power spectrum.  This will require calculating a different covariance matrix, $\mathbf{N}$, and computing $\mathbf{X}\Ninv\mathbf{X}^T$ per posterior calculation.  This latter matrix product can be computationally expensive due to the potentially large size of the matrix $\mathbf{X}$ with shape ($N_{\rm pix}, 2\envis$).  Due to these complications, it might still prove difficult to model all sky visibilities using the limited image-space model.

\subsection{Designing Instruments with Narrow Beams}

An alternative to modifying the model within \beor\ is modifying the instrument from which the data are obtained.  As it currently stands, \beor\ is particularly well suited to work with narrow primary beams.  The Gaussian beam used here, for example, performs well even in the presence of all sky FGs due to the suppression of sources far out in the beam.  A purely hardware solution would involve designing an instrument with a narrow beam and suppressed sidelobes.  Such instruments would be ideal candidates for use with our as-is analysis.

To test a beam with additional sidelobe suppression, we simulated all sky EoR+GSM+GLEAM datasets using a toy model beam comprised of an Airy beam multiplied by an achromatic Gaussian taper with a FWHM of 9.3$^\circ$.  While a Gaussian $\times$ Airy beam might not be a physically realizable beam shape, we use this toy model tapered beam here to show that reducing sidelobe amplitudes improves power spectrum recovery.  As mentioned in section \ref{sec:eor-fgs-all-sky}, the combination of an Airy beam with all sky FGs was numerically unstable due to the level of FG contamination in the EoR power spectrum.  When using the Tapered Airy beam with all sky FGs, power spectrum estimates consistent with expectation were obtained for all $k$.  Multiple FWHM values for the Gaussian taper were tried, but any FWHM $\gtrsim11^\circ$ produced FG contamination at low $k$.  This result also suggests that bright FGs just outside the model FoV can cause modelling problems.  This is consistent with the results for the Airy beam in \figref{fig:map-gsm-model-errors} which shows an increase in $\chi_k^2$ when including FGs that extend $\sim10^\circ$ beyond the model FoV (relative to the Gaussian line which serves as a reference for a ``working'' $\chi_k^2$ value).  Because FGs are so bright relative to the EoR, relatively small FG modelling errors can lead to significant EoR power spectrum contamination.  When using a sky model restricted to low zenith angles, the beam must significantly downweight FGs outside the model FoV to ensure unbiased power spectrum recovery.

For comparison, there are existing and next generation experiments with sufficiently narrow beams which might be suitable candidates for use with the as-is analysis.  The high band array (HBA, 120-240 MHz) of LoFAR, an existing experiment, might be one such candidate.  Simulated \citep{lofar} and observed \citep{lofar-beams} HBA station beams have FWHM values on the order of $\sim2-4^\circ$ at the frequencies used here \citep{lofar, lofar-beams}.  These beams are much narrower than the aforementioned toy model beam, however, the amplitude of the high zenith angle sidelobes is comparable to the Airy beam used here ($\sim30-40$ dB down from the peak).  Alternatively, the Square Kilometer Array (SKA, \citet{SKA}) provides a point of comparison for a next generation 21 cm interferometer.  Simulated SKA low frequency array (SKA-LOW) station beams from \citet{ska-low-beam-2016} and \citet{ska-low-beam-2017} are narrower than the Airy beam used here but have comparable sidelobes down only $\sim$40 dB from the peak.  However, \citet{ska-low-beam-2016} demonstrate that a 10 dB reduction in the amplitude of the first sidelobe of an SKA-LOW station beam can be obtained with a small loss in array sensitivity (15\%) via Taylor apodization.  The same approach could be used with an increased level of apodization to suppress station sidelobes at the required level for power spectral estimation of SKA-LOW data with BayesEoR.

\section{Conclusions}\label{sec:conclusion}

We have presented an updated version of our Bayesian power spectrum estimation pipeline for 21 cm interferometric data, \beor.  We performed a suite of tests detailing the performance of the code in its updated form on simulated visibilities, demonstrating the impact of incomplete FG modelling during power spectrum estimation.  The simulated datasets were generated for mock EoR only and mock EoR + FG sky models using either subsets of (restricted FoV) or the entirety of (all sky) the sky.  For each sky model, we simulated a dataset using a Gaussian and Airy beam.  In the absence of FGs, \beor\ is capable of producing power spectrum estimates consistent with expectation at all $k$ in the model for both restricted FoV and all sky datasets and both beam types.  When using EoR + FGs in the restricted FoV scenario, \beor\ again recovered power spectrum estimates consistent with expectation.  When using all sky EoR + FGs, only the Gaussian beam dataset was capable of unbiased power spectrum recovery at all $k$.  The combination of Airy beam with all sky EoR + FGs, however, produced severely contaminated EoR power spectrum estimates at all $k$.  The amplitudes of the sidelobes in the Airy beam were shown to insufficiently downweight bright FGs that lie outside of the extent of the image-space model.  These high zenith angle FGs contribute delays and fringe rates to the visibilities that the limited image-space model is incapable of recreating.

We also proposed several techniques that can be employed to improve our code.  Increasing the FoV of the joint EoR + FG model to model the entire sky is in principle possible, but it requires far too much memory for even modern supercomputers.  Separating the EoR and FG models to have their own individual fields of view is the most promising approach for improving our code and is left as future work.
Modifying the likelihood to perform a comparison of data and model in the image domain is another potentially interesting approach, but its efficacy requires further investigation.
Aside from improving the model in \beor, we also presented two ways in which our recovered power spectrum estimates could be improved.  Improving our understanding of the FG signals on the sky could allow us to first subtract a model of high zenith angle emission and fit a power spectrum to the residual visibilities (see section \ref{sec:fg-mitigation-egsm} for a discussion regarding the difference between our suggested approach and those taken by the MWA \citep{mwa-limits} and LOFAR \citep{lofar-limits}).  We also showed that a telescope with a realistic primary beam (Airy disk) but lower amplitude sidelobes would also allow for better power spectrum recovery without the need for increased compute resources.

Bright FGs pose great problems for current 21 cm power spectrum analysis pipelines.  Code that is capable of accounting for the covariance between the EoR and FG signals is essential to produce statistically rigorous EoR power spectrum estimates.  As the 21 cm cosmology community progresses toward a detection and characterization of the EoR, approaches similar to that described here will be vital for obtaining unbiased estimates of the power spectrum.

\section*{Acknowledgements}

The authors acknowledge support from NSF Awards \#1636646 and \#1907777, as well as Brown University's Richard B. Salomon Faculty Research Award Fund. JB also acknowledges support from a NASA RI Space Grant Graduate Fellowship. PHS was supported in part by a McGill Space Institute fellowship and funding from the Canada 150 Research Chairs Program.  This result is part of a project that has received funding from the European Research Council (ERC) under the European Union's Horizon 2020 research and innovation programme (Grant agreement No. 948764; JTB)  This research was conducted using computational resources and services at the Center for Computation and Visualization, Brown University.

%%%%%%%%%%%%%%%%%%%%%%%%%%%%%%%%%%%%%%%%%%%%%%%%%%
\section*{Data Availability}

The data used here will be provided upon reasonable request to the corresponding author.



\bibliographystyle{mnras}
\bibliography{bibliography}

%\begin{thebibliography}{99}
%\end{thebibliography}

%%%%%%%%%%%%%%%%%%%%%%%%%%%%%%%%%%%%%%%%%%%%%%%%%%

%%%%%%%%%%%%%%%%% APPENDICES %%%%%%%%%%%%%%%%%%%%%

% \appendix

%%%%%%%%%%%%%%%%%%%%%%%%%%%%%%%%%%%%%%%%%%%%%%%%%%

% Don't change these lines
\bsp	% typesetting comment
\label{lastpage}
\end{document}